\documentclass[aps,prd,groupaddress,usenatbib,onecolumn,nofootinbib,altaffilletter]{revtex4}
\usepackage{amsmath,amssymb}
\usepackage[utf8]{inputenc}
\usepackage{hyperref}
\usepackage{graphicx}
\usepackage{subfigure}

\def\be{\begin{equation}}
\def\ee{\end{equation}}

\begin{document}

\title{Non-commutative inspired black holes in Euler-Heisenberg non-linear electrodynamics}

\author{Marco Maceda}
\affiliation{Departamento de F\'{\i}sica, Universidad Aut\'onoma Metropolitana-Iztapalapa,\\A.P. 55-534, Mexico D.F. 09340, M\'exico.}
\email{mmac@xanum.uam.mx}

\author{Alfredo Mac\'{\i}as}
\affiliation{Departamento de F\'{\i}sica, Universidad Aut\'onoma Metropolitana--Iztapalapa,\\A.P. 55-534, Mexico D.F. 09340, M\'exico.}
\email{amac@xanum.uam.mx}

\date{\today}

\begin{abstract}
We find non-commutative inspired electrically and magnetically charged black hole solutions in Euler-Heisenberg non-linear electrodynamics. For these solutions, we determine the non-commutative corrections to the horizon radius for the general and extremal case. We also analyse the weak, dominant and strong energy conditions and the shadow associated with these metrics.
\end{abstract}


\maketitle

\section{Introduction}
\label{secc:1}

A quantum theory of gravity is a central challenge nowadays. A certain number of proposals exist to analyse quantum effects in gravitational fields (loop quantum gravity, string theory, non-commutative geometry, matrix geometry). All of them cover several aspects at different levels and complement each other; this interplay provides us with useful insights into the whole picture.

In some of these approaches, the structure of space-time is assumed to lose its continuum character. Such discretisation implies generalised incertitude principles that are natural consequences of a quantum theory of gravity where we have a set of coordinate and momentum operators with a discrete spectrum. In the general case, the commutation relations among the coordinates and momentum operators imply the existence of a minimal length~\cite{Maggiore:1993rv}. This length serves as a natural cutoff that removes divergences from the theory.

Non-commutative geometry~\cite{Con94,Mad:1999,Muller:2008} is a formalism where commutation relations among coordinates and momenta operators find a fertile ground to flourish. They may be incorporated straightforwardly into a classical theory using different schemes~\cite{Connes:1996gi,Moffat:2000fv,Schupp:2009pt}. For example, if we want to implement commutation relations among spatial coordinates only, we may use the Moyal star product that is a consequence of the commutation relations and replaces the standard point-wise multiplication of functions.

The use of star products to encode non-commutative effects generally leads to perturbative calculations. More recently, an approach~\cite{Banerjee:2009xx} based on coherent states in non-commutative quantum mechanics allows the analysis of non-perturbative effects. This analysis, initially motivated by calculations in non-commutative quantum field theory~\cite{Smailagic:2003rp,Smailagic:2004yy}, showed that a non-commutative Gaussian smeared distribution is the appropriate replacement for the point-like behaviour of particles usually present in a commutative setup.

In General Relativity this idea allows the construction of non-commutative inspired black holes. These objects have the standard properties associated with black holes, but they are regular at the source of the gravitational field. Nowadays, we know a rich variety of non-commutative inspired black hole solutions. They include the non-commutative inspired Schwarzschild and Reissner-Nordström (RN) metrics~\cite{Nicolini:2005vd,Nicolini:2008aj}. More recently, the Kerr and Kerr-Newman solutions have been obtained using a modified Janis-Newman algorithm specially tailored for the non-commutative framework~\cite{Modesto:2010rv}.

On the other hand, non-linear electrodynamics extends our knowledge of the electromagnetic field and its physical effects. It is a natural consequence when looking for a solution to the self-energy problem of a point charged particle. The Born-Infeld (BI) electrodynamics~\cite{Born:1934gh} is the first example of this. It is also a natural outcome when taking into account loop corrections in QED for instance, where the Euler-Heisenberg (EH) electrodynamics~\cite{Heisenberg:1935qt} becomes relevant. Both electrodynamics describe phenomena outside the realm of standard Maxwell's equations.

The recent observations of Sgr A* indicate that a massive black hole lies at the centre of the Milky Way, and this feature is believed to be present in the majority of the active galactic nuclei known to date. We expect then modifications on the behaviour of particles at the vicinity of these supermassive black holes; in this regard, the orbital motion of photons is a useful tool to determine the shadow of the black hole~\cite{Grenzebach:2014fha,Grenzebach:2015uva}. We have then a testable ground for theories analysing the quantum structure of spacetime at microscopic length scales. Furthermore, since the medium around the black hole involves matter interacting not only gravitationally but also electromagnetically, the existence of jets of charged particles is a common phenomenon, we also expect effects due to non-linear electrodynamics to be present as well. Previous results in this direction may be found in the literature~\cite{Sharif:2016znp,Saha:2018zas}.

To gain more insight into the several aspects that arise in the above situation, we consider in this work non-commutative inspired charged black holes in EH electrodynamics. We obtain non-commutative effects by using smeared distributions of mass and charge, and we choose to work with EH non-linear electrodynamics because it contains all the characteristic features present in more complex non-linear Electrodynamics, such as BI electrodynamics. Furthermore, from a practical point of view, it is more amenable to give us analytic results.

We organise this paper as follows: in Sec.~\ref{secc:2} we review the electrically and magnetically charged black hole solutions in classical EH electrodynamics. We then construct the corresponding non-commutative inspired black holes in Sec.~\ref{secc:3}, and we calculate the corrections to the horizon radius of the non-commutative metrics in Sec.~\ref{secc:4}. We also analyse the strong, dominant and weak energy conditions in Sec.~\ref{secc:5} and we investigate the shadow of the non-commutative inspired black holes in Sec.~\ref{secc:6} . We end with some remarks and perspectives in the Conclusions.

\section{Static charged black holes in Euler-Heisenberg electrodynamics}
\label{secc:2}

EH electrodynamics is a low-energy limit of BI electrodynamics. Its Lagrangian is
\be
{\cal L}_{EH} = -x + \frac A2 x^2 + \frac B2 y^2,
\label{ehlag}
\ee
where
\be
x := \frac 14 F_{\mu\nu} F^{\mu\nu}, \qquad y := \frac 14 F_{\mu\nu} \star F^{\mu\nu},
\label{invars}
\ee
are the relativistic invariants of the electromagnetic field. Here $F_{\mu\nu}$ is the electromagnetic tensor.

The EH Lagrangian can be written using the Plebanski variables~\cite{Plebanski:1970,Gibbons:2000xe,Gibbons:2001sx}
\be
P_{\mu\nu} := -({\cal L}_x F_{\mu\nu} + {\cal L}_y \star F_{\mu\nu}),
\ee
and their Hodge duals $\star P_{\mu\nu}$. Then, the dual invariants are
\be
s := -\frac 14 P_{\mu\nu} P^{\mu\nu}, \qquad t := -\frac 14 P_{\mu\nu} \star P^{\mu\nu},
\label{dualinvars}
\ee
and we have the dual Hamiltonian
\be
\hat {\cal L}_{EH} = s - \frac A2 s^2 - \frac B2 t^2.
\label{dualehlag}
\ee
We should remark that in the dual description of the EH Lagrangian, we do not consider terms higher than linear for both parameters $A$ and $B$. On the other hand, the field equations of gravity coupled to a general nonlinear electrodynamics come from variations of the action~\cite{Gibbons:2001sx}
\be
S = \frac 1{16\pi G_N} \int d^4x \,\sqrt{-g}\,R + \frac 1{4\pi}\int d^4x \,\sqrt{-g}\, {\cal L}(x,y),
\ee
with respect to the metric and electromagnetic potentials. Explicitly, we have that the field equations are
\be
G_{\mu\nu} = 8\pi T_{\mu\nu}, \qquad \nabla_\mu P^{\mu\nu} = 0,
\ee
where the energy-momentum tensor is
\be
4\pi T_{\mu\nu} = \hat {\cal L}_s P_{\mu\alpha} g^{\alpha\beta} P_{\nu\beta} + g_{\mu\nu} (2s \hat {\cal L}_s + t \hat
{\cal L}_t - \hat {\cal L}).
\ee

Let us consider a spherically symmetric line element
\be
ds^2 = - \left( 1 - \frac {2m(r)}r \right) dt^2 + \left( 1 - \frac {2m(r)}r \right)^{-1} dr^2 + r^2 d\Omega,
\ee
where $d\Omega = d\vartheta^2 + \sin^2 \vartheta d\phi^2$. For the electric solution, the natural description is in terms of the Plebanski variables $P_{\mu\nu}$~\cite{Bronnikov:2000yz,Bronnikov:2000vy,Burinskii:2002pz,Bronnikov:2017tnz}; we have that
\be
P_{\mu\nu} = \frac {Q_e}{r^2} \, \delta^0_{[\mu} \delta^1_{\nu]},
\label{cpmunu}
\ee
fulfils the conservation laws $\nabla_\mu P^{\mu\nu} = 0$ in this situation. The relevant gravitational field equation is then
\be
\frac {m_{,r}}{r^2} = \frac 12 P_{01}^2 - \frac 18 A P_{01}^4,
\label{feque}
\ee
and its solution is given by
\be
m(r) = M - \frac {Q_e^2}{2r} + \frac A{40} \frac {Q_e^4}{r^5}.
\ee
The associated metric is then
\begin{eqnarray}
ds^2 &=& -\left( 1 - \frac {2M}r + \frac {Q_e^2}{r^2} - \frac A{20} \frac {Q_e^4}{r^6} \right) dt^2
\nonumber \\[4pt]
&&+ \left( 1 - \frac {2M}r + \frac {Q_e^2}{r^2} - \frac A{20} \frac {Q_e^4}{r^6} \right)^{-1} dr^2 + r^2 d\Omega.
\label{ecbh}
\end{eqnarray}
Here $M$ represents the mass of the point-like source at the origin.

On the other hand, the electromagnetic tensor $F_{\mu\nu}$ provides the natural description of the magnetic charged solution. In this case, we have that
\be
F_{\mu\nu} = - Q_m \cos \vartheta \, \delta^2_{[\mu} \delta^3_{\nu]},
\label{cfmunu}
\ee
satisfies the conservation laws and the relevant gravitational field equation is
\be
\frac {m_{,r}}{r^2} = - s + \frac {3A}2 s^2, \qquad s := -\frac 12 \frac {Q_m^2}{r^4} \left( -1 + \frac A2 \frac {Q_m^2}{r^4} \right)^2.
\label{fequm}
\ee
Its solution is given by
\be
m(r) = M - \frac {Q_m^2}{2r} + \frac A{40} \frac {Q_m^4}{r^5},
\ee
and the associated metric is then
\begin{eqnarray}
ds^2 &=& -\left( 1 - \frac {2M}r + \frac {Q_m^2}{r^2} - \frac A{20} \frac {Q_m^4}{r^6} \right) dt^2
\nonumber \\[4pt]
&&+ \left( 1 - \frac {2M}r + \frac {Q_m^2}{r^2} - \frac A{20} \frac {Q_m^4}{r^6} \right)^{-1} dr^2 + r^2 d\Omega.
\label{mcbh}
\end{eqnarray}
As previously, $M$ represents the mass of the point-like source at the origin. Notice that the magnetic charged solution has the same functional form as the electrical charged solution.

\begin{center}
\begin{figure}[h!]
\includegraphics[width=9cm]{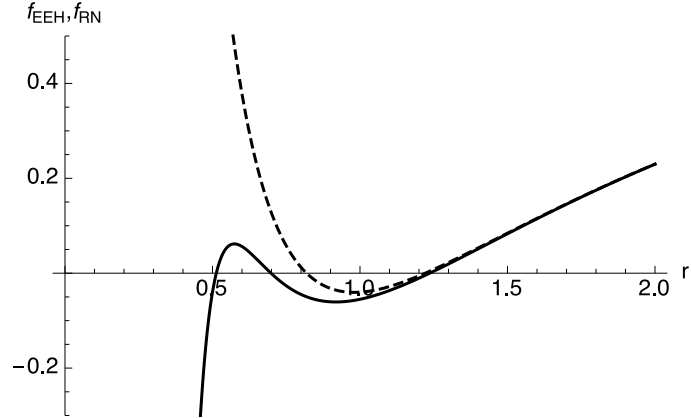}
\caption{Behaviour of $f(r)$ in EEH spacetime with $Q_e =1, A = 0.3$ (solid line) and $f(r)$ in RN (dashed line) with $Q_e =1$. The three distinct horizons for Einstein-Euler-Heisenberg are located at $r = 0.5129, 0.698322,1.23638$ and those of Reissner-Nordström at $r = 0.819002, 1.221$. In both cases $M = 1.02 > Q_e$.}
\label{graph00}
\end{figure}
\end{center}

Each one of the spacetime metrics given previously has horizons; they satisfy the condition $f(r_0) = 1 -2 m(r_0)/r_0 = 0$. An exact expression for the horizon is difficult to obtain, nevertheless if we look for a perturbative solution of the form
\begin{eqnarray}
r_0^{-1} = \beta + A\gamma,
\end{eqnarray}
where we assume $A$ to be small, then we have the following two answers
\begin{widetext}
\begin{eqnarray}
\beta_+ &=& \frac 1{M+\sqrt{M^2-Q_e^2}},
\nonumber \\[4pt]
\gamma_+ &=& \frac{\frac{32 M^7}{Q_e^2}-64 M^5+38 M^3 Q_e^2-18 M^2 Q_e^2 \sqrt{M^2-Q_e^2}+Q_e^4
   \sqrt{M^2-Q_e^2}-\frac{32 M^6 \sqrt{M^2-Q_e^2}}{Q_e^2}+48 M^4 \sqrt{M^2-Q_e^2}-6 M Q_e^4}{40
   \left(M^2 -Q_e^2\right) Q_e^6},
\label{classh1p}
\end{eqnarray}
and
\begin{eqnarray}
\beta_- &=& \frac 1{M-\sqrt{M^2-Q_e^2}},
\nonumber \\[4pt]
\gamma_- &=& \frac{\frac{32 M^7}{Q_e^2}-64 M^5+38 M^3 Q_e^2+18 M^2 Q_e^2 \sqrt{M^2-Q_e^2}-Q_e^4
   \sqrt{M^2-Q_e^2}+\frac{32 M^6 \sqrt{M^2-Q_e^2}}{Q_e^2}-48 M^4 \sqrt{M^2-Q_e^2}-6 M Q_e^4}{40(M^2-Q_e^2) Q_e^6}.
\label{classh1m}
\end{eqnarray}
\end{widetext}
We need $M \geq Q_e$ for these values to be real. The subscript plus or minus refers to the distinct values of the outer and inner horizons of the Reissner-Nordström limit ($A=0$). In Fig.~\ref{graph00} we show this generic situation where the outer horizons in EEH and RN almost coincide.

As we know, RN spacetime has an extremal case defined by the conditions $f(r_e) = 0, f_{,r}(r_e) = 0$. The analogous extremal configuration in EEH spacetime appears when the extremal horizon radius $r_e$ satisfies the conditions
\begin{eqnarray}
0 &=& 1 - \frac {2M}{r_e} + \frac {Q_e^2}{r_e^2} - \frac A{20} \frac {Q_e^4}{r_e^6},
\nonumber \\[4pt]
0 &=& \frac {2M}{r_e^2} - 2 \frac {Q_e^2}{r_e^3} + \frac {3A}{10} \frac {Q_e^4}{r_e^7}.
\end{eqnarray}
It follows that
\be
1 - \frac {Q_e^2}{r_e^2} + \frac A4 \frac {Q_e^4}{r_e^6} = 0.
\ee
For a given value of $Q_e$ and $A$, the previous equation provides a horizon radius $r_e$ that in turn gives a value for the extremal mass using
\be
M_e = -\frac 3{20} \frac {AQ_e^4}{r_e^5} + \frac {Q_e^2}{r_e}.
\ee

In Fig.~\ref{fa}, we show again a situation with three distinct horizons. From it, if we treat the Euler-Heisenberg contribution as a perturbation, there are two different extremal cases to consider:
\begin{enumerate}
\item The two innermost horizons become a single one: The extremal horizon radius is
\be
r_e = \sqrt{\frac {Q_e}2} A^{1/4} + \frac {A^{3/4}}{8\sqrt{2Q_e}}  + \frac {9A^{5/4}}{128 \sqrt{2}Q_e^{3/2}},
\ee
and the relation between charge and mass is
\begin{eqnarray}
M_e &=& \frac {2\sqrt{2} Q_e^{3/2}}{5A^{1/4}} + \frac {\sqrt{Q_e} A^{1/4} }{2\sqrt{2}}
\nonumber \\[4pt]
&&+ \frac {A^{3/4}}{32\sqrt{2Q_e}} + \frac {3A^{5/4}}{256\sqrt{2}Q_e^{3/2}}.
\end{eqnarray}
This situation is well illustrated in Fig.~\ref{fb}. Notice the appearance of fractional powers of $A$ in the expressions for the mass and the horizon.

\begin{figure}[h!]
\centering
\subfigure[] 
{
    \label{fa}
    \includegraphics[width=0.4\textwidth]{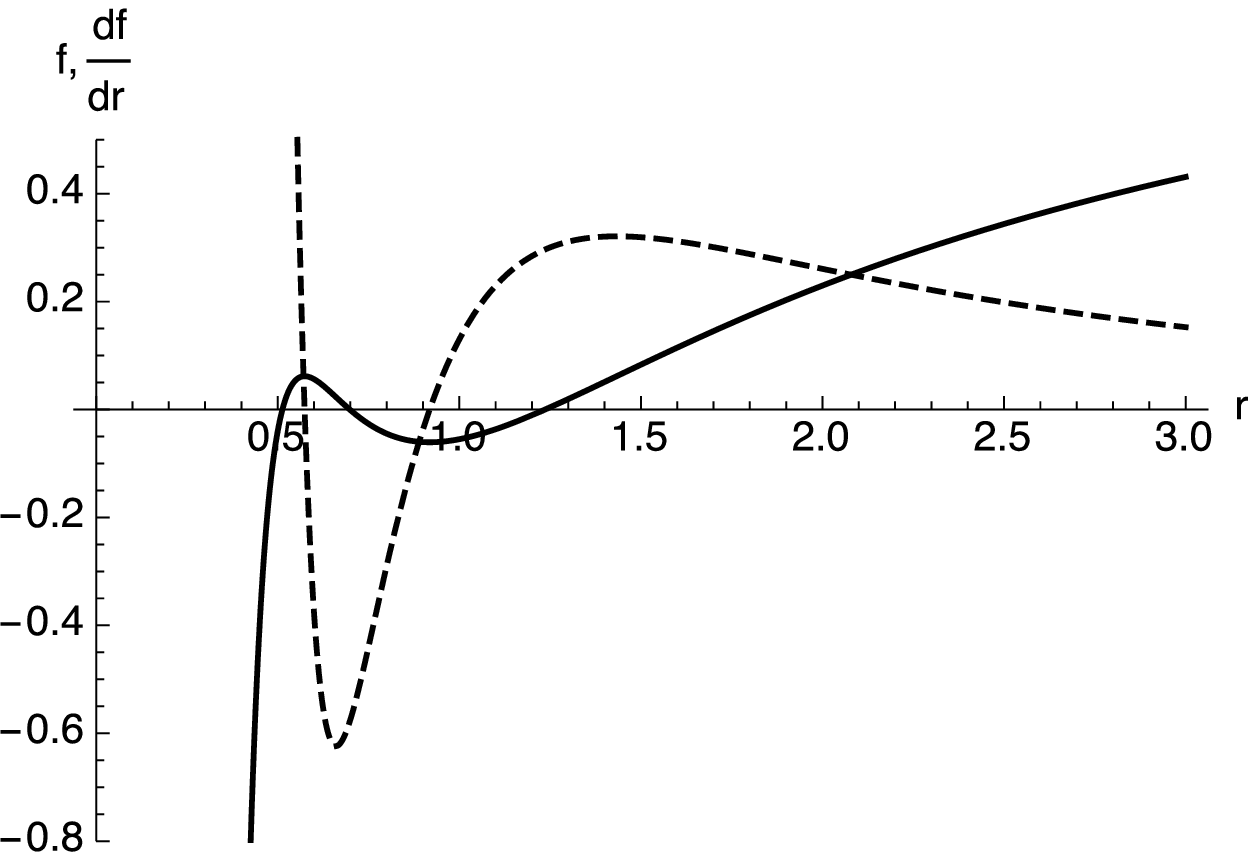}
}
\subfigure[] 
{
    \label{fb}
    \includegraphics[width=0.4\textwidth]{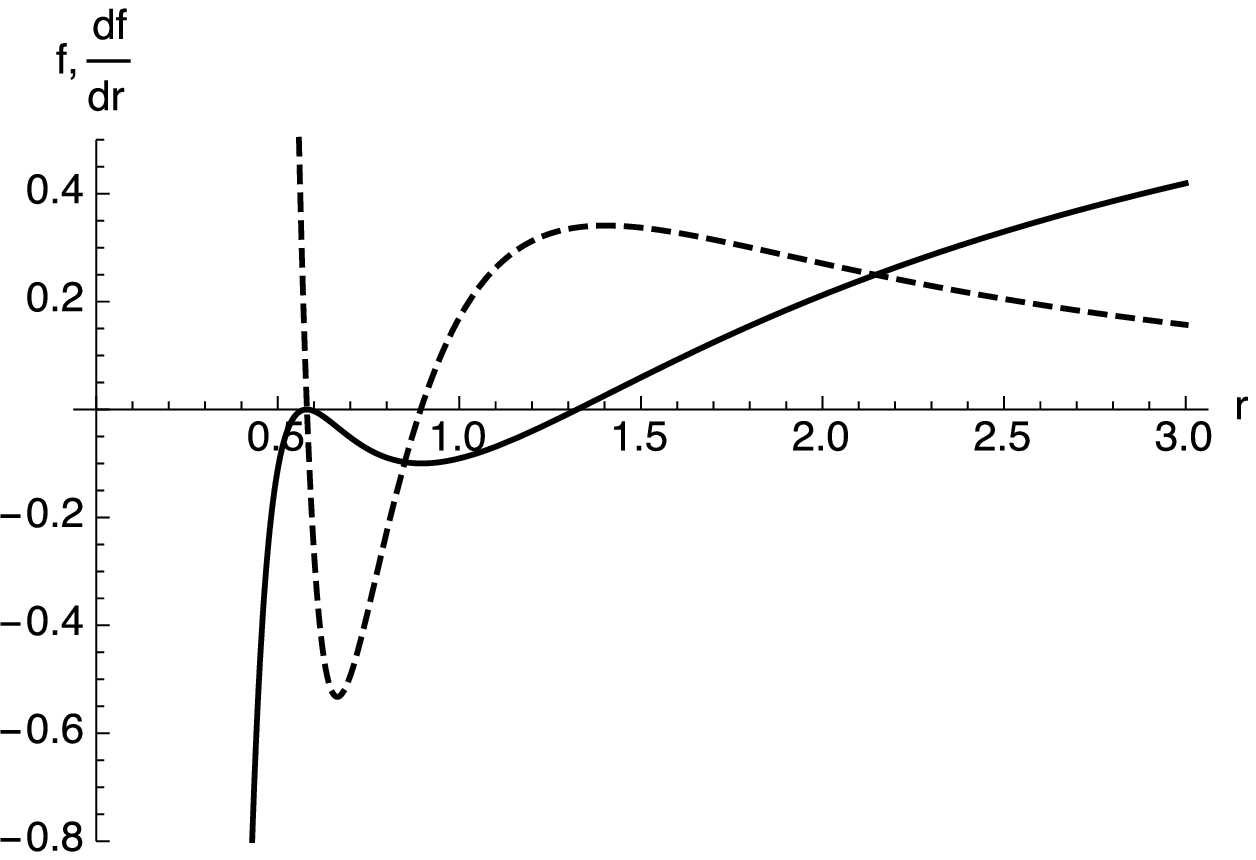}
}
\subfigure[] 
{
    \label{fc}
    \includegraphics[width=0.4\textwidth]{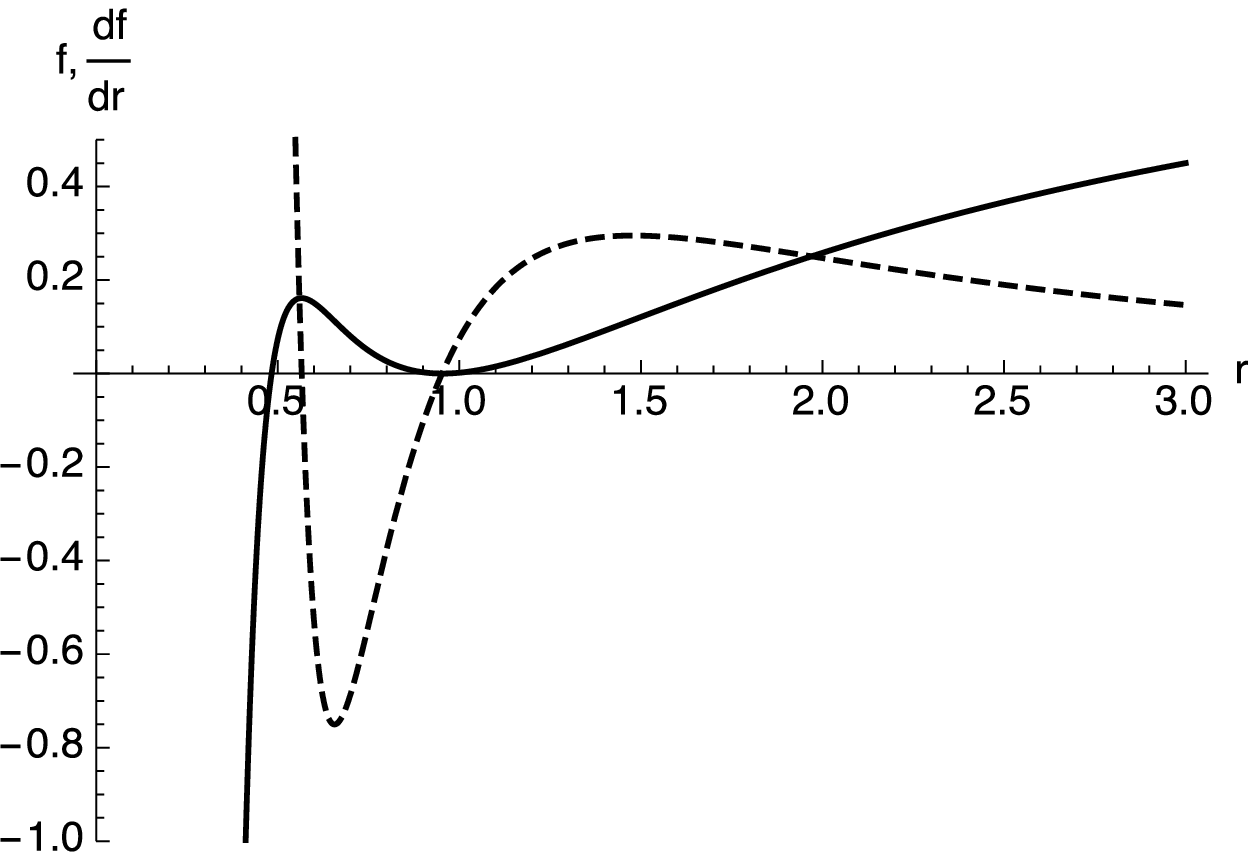}
}
\caption{Behaviour of $f(r)$ (solid line) and $f_{,r}(r)$ (dashed line) in EEH spacetime for $Q_e =1, A = 0.3$. In Fig.~\ref{fa}, three distinct horizons exist; they are located at $r = 0.5129, 0.698322,1.23638$ with $M = 1.02 > Q_e$.  In Fig.~\ref{fb} the two inner horizons become a single inner horizon located at $r_e = 0.579751$ with $M_e = 1.0378 > Q_e$. In Fig.~\ref{fc}, the two outer horizons become a single exterior horizon located at $r_e = 0.953567$ with $M_e = 0.991618 < Q_e$.}
\end{figure}

\item The two outermost horizons become a single one: The extremal horizon radius is
\be
r_e = Q_e \left( 1 - \frac A8 Q_e^{-2} \right),
\label{re2}
\ee
and the relation between mass and charge is
\be
M_e = Q_e \left( 1 - \frac A{40} Q_e^{-2} \right).
\label{me2}
\ee
As seen from this expression, mass {\it has} a lower value than charge; this situation is illustrated in Fig.~\ref{fc}. In Tables~\ref{table:1} and~\ref{table:2} we show, for different values of the charge, some numerical values for $r_e$ and $M_e$ using Eqs.~(\ref{re2}) and~(\ref{me2}). We see that the approximate solution is in good agreement with the exact result.

\begin{table}[h!]
\centering
\caption{Comparison between the values of the extremal radius and mass using the approximate solution and the exact numerical result.\label{table:1}}
\begin{tabular}{ |p{1cm}||p{2cm}|p{2cm}|p{2cm}|p{2cm}|}
 \hline
 \multicolumn{5}{|c|}{$Q=1$} \\
 \hline
 $A$ & $r_e$ (exact) & $r_e$ (approx) & $M_e$ (exact) & $M_e$ (approx) \\
 \hline
 0.1& 0.986726 & 0.9875 & 0.997416 & 0.9975\\
 \hline
 0.2 &0.971533 & 0.975 & 0.994641& 0.995\\
 \hline
 0.3 & 0.953567 & 0.9625 & 0.991618 & 0.9925\\
 \hline
 0.4 & 0.931102 & 0.95 & 0.98826 & 0.99\\
 \hline
 0.5 & 0.899454 & 0.9375 & 0.984386 & 0.9875\\
 \hline
\end{tabular}
\end{table}

\begin{table}[h!]
\centering
\caption{Comparison between the values of the extremal radius and mass using the approximate solution and the exact numerical result.\label{table:2}}
\begin{tabular}{ |p{1cm}||p{2cm}|p{2cm}|p{2cm}|p{2cm}|}
 \hline
 \multicolumn{5}{|c|}{$Q=2$} \\
 \hline
 $A$ & $r_e$ (exact) & $r_e$ (approx) & $M_e$ (exact) & $M_e$ (approx) \\
 \hline
 0.1& 1.99366 & 1.99375 & 1.99874 & 1.99875\\
 \hline
 0.2 & 1.98713 & 1.9875 & 1.99746 & 1.9975\\
 \hline
 0.3 & 1.9804 & 1.98125 & 1.99616 & 1.99625\\
 \hline
 0.4 & 1.97345 & 1.975 & 1.99483 & 1.995\\
 \hline
 0.5 & 1.96627 & 1.96875 & 1.99348 & 1.99375\\
 \hline
 0.6& 1.95882 & 1.9625 & 1.99211 & 1.9925\\
 \hline
 0.7 & 1.9511 & 1.95625 & 1.99071 & 1.99125\\
 \hline
 0.8 & 1.94307 & 1.95 & 1.98928 & 1.99\\
 \hline
 0.9 & 1.9347 & 1.94375 & 1.98782 & 1.98875\\
 \hline
\end{tabular}
\end{table}

\end{enumerate}

\section{Non-commutative  inspired black holes with electric and magnetic charge from Euler-Heisenberg electrodynamics}
\label{secc:3}

Non-commutative inspired models are solutions of the modified gravitational field equations~\cite{Smailagic:2003rp,Smailagic:2004yy,Banerjee:2009xx,Nicolini:2005vd,Nicolini:2008aj,Tejeiro:2010gu,Liang:2012vx,Rahaman:2013gw,Gonzalez:2014mza}
\be
R_{\mu\nu} - \frac 12 g_{\mu\nu} R = 8\pi [\mathfrak T_{\mu\nu} + T_{\mu\nu}^{e.m.}],
\ee
where
\be
\mathfrak T^\mu{}_{\nu} := diag (h_1, h_1, h_3, \dots, h_3), \quad h_3 := (r^2 h_1)_{,r}/2r,
\ee
is a non-commutative energy-matter tensor and $T_{\mu\nu}^{e.m.}$ is the standard electromagnetic tensor. The function $h_1$ in $\mathfrak T^\mu{}_{\nu}$ is given by
\be
h_1 := -\rho_m(r) = -\frac {\cal M}{(4\pi\theta)^{3/2}} e^{-r^2/4\theta},
\ee
where $\theta$ is the non-commutative parameter and $h_3$ is defined such that the conservation law $\nabla_\mu \mathfrak T^{\mu\nu} = 0$ holds; the normalization of $\rho_m(r)$ is such that
\be
\int d^d x \, \rho_m (r) = {\cal M},
\ee
gives the mass of a classical point-like source of the gravitational field. In the commutative limit $\theta \to 0$, the smeared distribution $\rho_m (r)$ becomes a Dirac delta function. The motivation and justification of these smeared distributions to encode non-commutative effects where first discussed in the context of quantum field theory and afterwards extended to the gravitational arena in the static and rotating scenarios~\cite{Smailagic:2003rp,Smailagic:2004yy,Banerjee:2009xx,Nicolini:2005vd,Gonzalez:2014mza,Modesto:2010rv}.

\subsection{Non-commutative inspired electric solution}

In this section, we give the non-commutative inspired counterparts of the solutions discussed previously. First, for the electrically charged case we solve the non-commutative conservation laws
\be
\nabla_\mu P^{\mu\nu} = 4\pi J^\mu,
\label{ncef}
\ee
where
\be
J^{\mu} = Q_e \left[ \frac {e^{-r^2/4\theta}}{(4\pi \theta)^{3/2}}, 0, 0 , 0 \right],
\ee
is a source for the electric field having a non-commutative origin. This source basically replaces the point-like behaviour of the delta function by a electrically charged smeared distribution depending on the non-commutative parameter.

The solution to Eq.~(\ref{ncef}) is then given by the non-commutative inspired Plebanski variables
\be
P_{\mu\nu} = \frac 2{\sqrt{\pi}} \frac {Q_e}{r^2} \gamma \left( \frac 32, \frac {r^2}{4\theta} \right) \, \delta^0_{[\mu} \delta^1_{\nu]},
\label{ncp01}
\ee
where $\gamma (a, z)$ is the lower incomplete gamma function~\cite{Abramowitz:1965}. Notice that in the commutative limit $\theta \to 0$, we recover the commutative result in Eq.~(\ref{cpmunu}) since $\lim_{\theta \to 0} \gamma \left( \frac 32, \frac {r^2}{4\theta} \right) = \frac {\sqrt{\pi}}2$. Similar to the commutative case, the relevant field equation is now
\be
\frac {m_{,r}}{r^2} = \frac 12 P_{01}^2 - \frac 18 A P_{01}^2 + \frac {\cal M}{2\sqrt{\pi} \theta^{3/2}} e^{-r^2/4\theta},
\label{ncfeque}
\ee
and its solution is then
\begin{eqnarray}
m(r) &=& \frac {2 \cal M}{\sqrt{\pi}} \gamma\left( \frac 32, \frac {r^2}{4\theta} \right) + \frac 2\pi Q_e^2 \int_0^r \frac {ds}{s^2} \gamma^2 \left( \frac 32, \frac {s^2}{4\theta} \right)
\nonumber \\[4pt]
&& - \frac {2 A}{\pi^2} Q_e^4 \int_0^r \frac {ds}{s^6} \gamma^4 \left( \frac 32, \frac {s^2}{4\theta} \right),
\end{eqnarray}
where $\cal M$ is the ``bare" mass. The non-commutative inspired electrically charged black hole in this case is then
\begin{eqnarray}
ds^2 &=& -\left[ 1 - \frac {4 \cal M}{r\sqrt{\pi}} \gamma\left( \frac 32, \frac {r^2}{4\theta} \right) - \frac 4\pi \frac {Q_e^2}r \int_0^r \frac {ds}{s^2} \gamma^2 \left( \frac 32, \frac {s^2}{4\theta} \right) \right.
\nonumber \\[4pt]
&& \left. + \frac {4 A}{\pi^2} \frac {Q_e^4}r \int_0^r \frac {ds}{s^6} \gamma^4 \left( \frac 32, \frac {s^2}{4\theta} \right) \right] dt^2
\nonumber \\[4pt]
&& + \left[ 1 - \frac {4 \cal M}{r\sqrt{\pi}} \gamma\left( \frac 32, \frac {r^2}{4\theta} \right) - \frac 4\pi \frac {Q_e^2}r \int_0^r \frac {ds}{s^2} \gamma^2 \left( \frac 32, \frac {s^2}{4\theta} \right) \right.
\nonumber \\[4pt]
&& \left. + \frac {4 A}{\pi^2} \frac {Q_e^4}r \int_0^r \frac {ds}{s^6} \gamma^4 \left( \frac 32, \frac {s^2}{4\theta} \right) \right]^{-1} dr^2
+ r^2 d\Omega.
\label{ncecbh}
\end{eqnarray}
This expression is more involved due to the appearance of the incomplete gamma function under the integrals, nevertheless one of them can be calculated as shown in the Appendix. Using this result and the ADM mass~\cite{Nicolini:2008aj}
\be
M := \oint_\Sigma d\sigma^\mu (T^0_\mu |_{matt} + T^0_\mu|_{el}),
\label{admmass}
\ee
we have
\begin{eqnarray}
ds^2 &=& - f(r) dt^2 + f(r)^{-1} dr^2 + r^2 d\Omega,
\label{ncecbhadm}
\end{eqnarray}
with
\begin{eqnarray}
f(r) &=& 1 - \frac {4M}{r\sqrt{\pi}} \gamma\left( \frac 32, \frac {r^2}{4\theta} \right) + \sqrt{\frac 2\theta} \frac 1\pi \frac {Q_e^2}r \gamma \left( \frac 32, \frac {r^2}{4\theta} \right)
\nonumber \\[4pt]
&&+ \frac 1\pi \frac {Q_e^2}{r^2} \left[ \gamma^2 \left( \frac 12, \frac {r^2}{4\theta} \right) - \frac r{\sqrt{2\theta}} \gamma \left( \frac 12, \frac {r^2}{2\theta} \right) \right]
\nonumber \\[4pt]
&&
- \frac {4 A}{\pi^2} \frac {Q_e^4}r \int_r^\infty \frac {ds}{s^6} \gamma^4 \left( \frac 32, \frac {s^2}{4\theta} \right)
\nonumber \\[4pt]
&&+ \frac {4 A}{\pi^2} \frac {Q_e^4}r \left[ 1- \frac 2{\sqrt{\pi}} \gamma \left( \frac 32, \frac {r^2}{4\theta} \right) \right] \int_0^\infty \frac {ds}{s^6} \gamma^4 \left( \frac 32, \frac {s^2}{4\theta} \right)
\nonumber \\[4pt]
&=& 1 - \frac {4M}{r\sqrt{\pi}} \gamma\left( \frac 32, \frac {r^2}{4\theta} \right) + \frac 1\pi \frac {Q_e^2}{r^2} \gamma^2 \left( \frac 12, \frac {r^2}{4\theta} \right)
\nonumber \\[4pt]
&& + \frac 1\pi \frac {Q_e^2}{r^2} \left[ \sqrt{\frac 2\theta} r \gamma \left( \frac 32, \frac {r^2}{4\theta} \right) - \frac r{\sqrt{2\theta}} \gamma \left( \frac 12, \frac {r^2}{2\theta} \right)  \right] 
\nonumber \\[4pt]
&&- \frac {4A}{\pi^2} \frac {Q_e^4}r \int_r^\infty \frac {ds}{s^6} \gamma^4 \left( \frac 32, \frac {s^2}{4\theta} \right)
\nonumber \\[4pt]
&&+ \frac {A}{8\pi^2} \frac {Q_e^4}r \left[ 1- \frac 2{\sqrt{\pi}} \gamma \left( \frac 32, \frac {r^2}{4\theta} \right) \right] \frac \alpha{\theta^{5/2}},
\label{dsnceeh}
\end{eqnarray}
where $\alpha := \int_0^\infty ds \,s^{-6} \gamma^4 \left( \frac 32, s^2 \right) = 0.02757$ and we have used the identity~\cite{Abramowitz:1965} $\gamma(\frac a2 + 1, z^2) = \frac a2 \gamma (\frac a2, z^2) - z^a e^{-z^2}$ to obtain the second equality. It is straightforward to verify that in the commutative limit $\theta \to 0$, we recover Eq.~(\ref{ecbh}).

\subsection{Non-commutative inspired magnetic solution}

The magnetic charged solution is defined by the non-commutative inspired electromagnetic tensor
\be
F_{\mu\nu} = - \frac 2{\sqrt{\pi}} Q_m \gamma \left( \frac 32, \frac {r^2}{4\theta} \right) \sin \vartheta \, \delta^2_{[\mu} \delta^3_{\nu]},
\label{ncfmunu}
\ee
and the field equation to be considered now is then
\be
\frac {m_{,r}}{r^2} = - s + \frac {3A}2 s^2 + \frac {\cal M}{2\sqrt{\pi} \theta^{3/2}} e^{-r^2/4\theta},
\label{ncfequm}
\ee
where $s$ is defined as in Eq.~(\ref{fequm}). Therefore, we have
\begin{eqnarray}
m(r) &=& \frac {2 \cal M}{\sqrt{\pi}} \gamma\left( \frac 32, \frac {r^2}{4\theta} \right) + \frac 2\pi Q_m^2 \int_0^r \frac {ds}{s^2} \gamma^2 \left( \frac 32, \frac {s^2}{4\theta} \right)
\nonumber \\[4pt]
&& - \frac {2 A}{\pi^2} Q_m^4 \int_0^r \frac {ds}{s^6} \gamma^4 \left( \frac 32, \frac {s^2}{4\theta} \right).
\end{eqnarray}
The non-commutative inspired magnetically charged black hole is
\begin{eqnarray}
ds^2 &=& -\left[ 1 - \frac {4 \cal M}{r\sqrt{\pi}} \gamma\left( \frac 32, \frac {r^2}{4\theta} \right) - \frac 4\pi \frac {Q_m^2}r \int_0^r \frac {ds}{s^2} \gamma^2 \left( \frac 32, \frac {s^2}{4\theta} \right) \right.
\nonumber \\[4pt]
&& \left. + \frac {4 A}{\pi^2} \frac {Q_m^4}r \int_0^r \frac {ds}{s^6} \gamma^4 \left( \frac 32, \frac {s^2}{4\theta} \right) \right] dt^2
\nonumber \\[4pt]
&& + \left[ 1 - \frac {4 \cal M}{r\sqrt{\pi}} \gamma\left( \frac 32, \frac {r^2}{4\theta} \right) - \frac 4\pi \frac {Q_m^2}r \int_0^r \frac {ds}{s^2} \gamma^2 \left( \frac 32, \frac {s^2}{4\theta} \right) \right.
\nonumber \\[4pt]
&& \left. + \frac {4 A}{\pi^2} \frac {Q_m^4}r \int_0^r \frac {ds}{s^6} \gamma^4 \left( \frac 32, \frac {s^2}{4\theta} \right) \right]^{-1} dr^2
+ r^2 d\Omega.
\label{ncmcbh}
\end{eqnarray}
As previously, this metric can be rewritten in terms of the ADM mass $M$ as defined in Eq.~(\ref{admmass}); we obtain then a metric similar to Eq.~(\ref{ncecbhadm}) but with $Q_e$ replaced by $Q_m$. Explicitly we have
\begin{eqnarray}
ds^2 &=& - f(r) dt^2 + f(r)^{-1} dr^2 + r^2 d\Omega,
\label{ncmcbhadm}
\end{eqnarray}
where
\begin{eqnarray}
f(r) &=& 1 - \frac {4M}{r\sqrt{\pi}} \gamma\left( \frac 32, \frac {r^2}{4\theta} \right) + \frac 1\pi \frac {Q_m^2}{r^2} \gamma^2 \left( \frac 12, \frac {r^2}{4\theta} \right)
\nonumber \\[4pt]
&& + \frac 1\pi \frac {Q_m^2}{r^2} \left[ \sqrt{\frac 2\theta} r \gamma \left( \frac 32, \frac {r^2}{4\theta} \right) - \frac r{\sqrt{2\theta}} \gamma \left( \frac 12, \frac {r^2}{2\theta} \right)  \right] 
\nonumber \\[4pt]
&&-\frac {4A}{\pi^2} \frac {Q_m^4}r \int_r^\infty \frac {ds}{s^6} \gamma^4 \left( \frac 32, \frac {s^2}{4\theta} \right)
\nonumber \\[4pt]
&&+ \frac {A}{8\pi^2} \frac {Q_m^4}r \left[ 1- \frac 2{\sqrt{\pi}} \gamma \left( \frac 32, \frac {r^2}{4\theta} \right) \right] \frac \alpha{\theta^{5/2}},
\end{eqnarray}
where $\alpha = 0.02757$. Using this last form, in the limit $\theta \to 0$ we recover Eq.~(\ref{mcbh}). The metrics in Eqs.~(\ref{ncecbh}) and~(\ref{ncmcbh}) are related by the same functional expression as it happens in the classical case.

\section{Horizon radius for the non-commutative Einstein-Euler-Heisenberg spacetime}
\label{secc:4}

\begin{figure}[h!]
\centering
\subfigure[] 
{
    \label{orbits:aa}
    \includegraphics[width=0.4\textwidth]{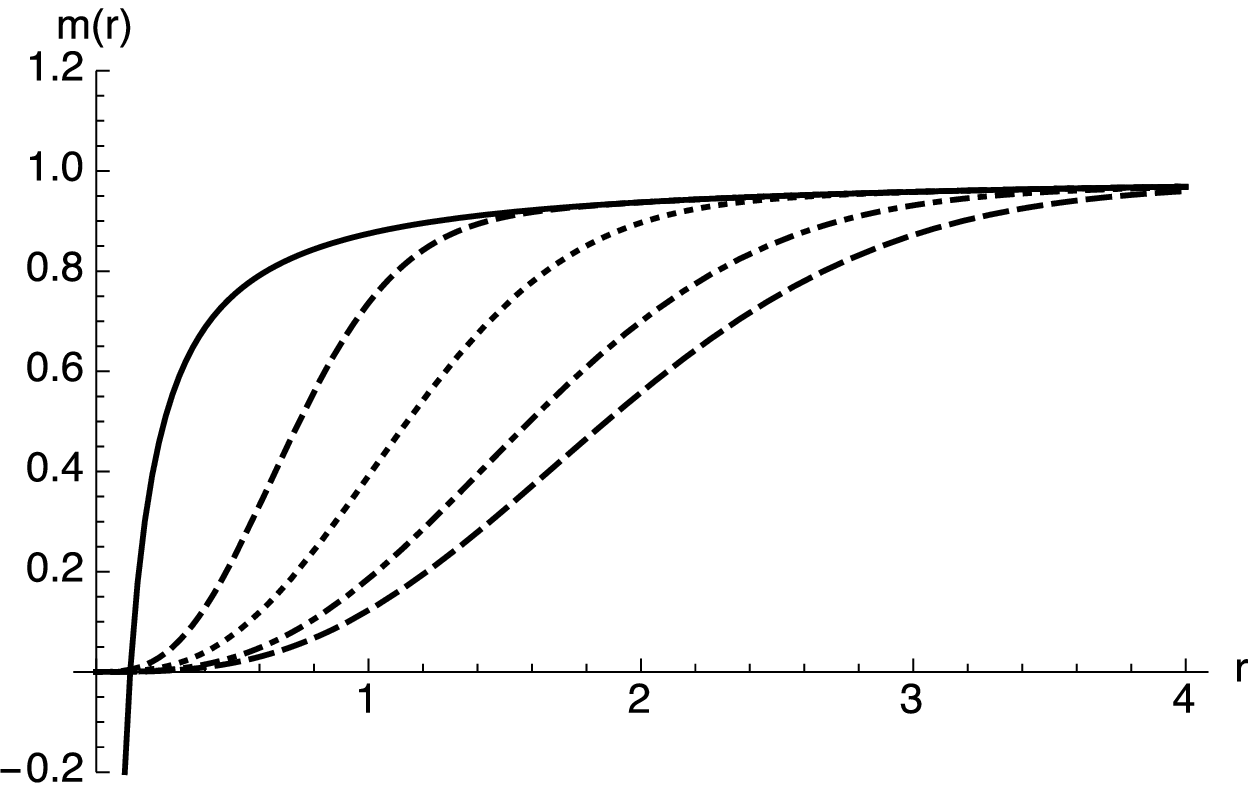}
}
\subfigure[] 
{
    \label{orbits:bb}
    \includegraphics[width=0.4\textwidth]{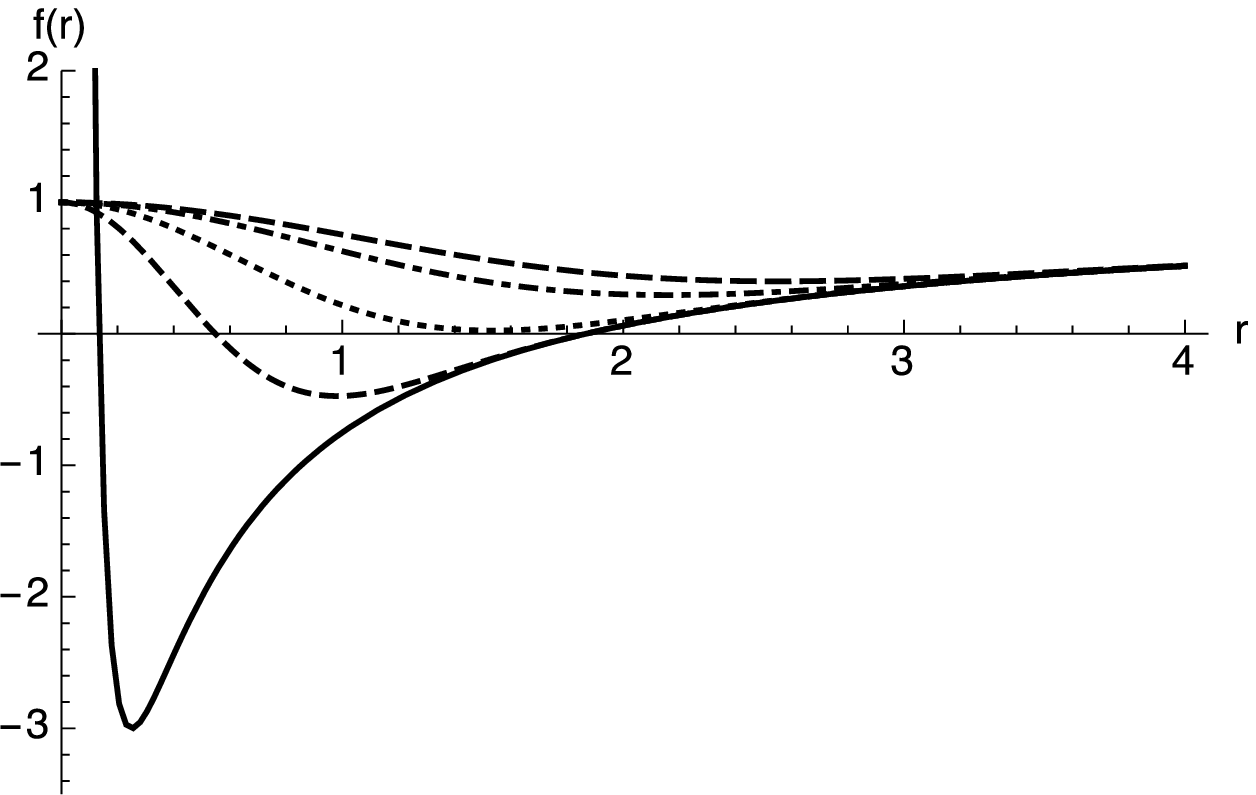}
}
\subfigure[] 
{
    \label{orbits:cc}
    \includegraphics[width=0.4\textwidth]{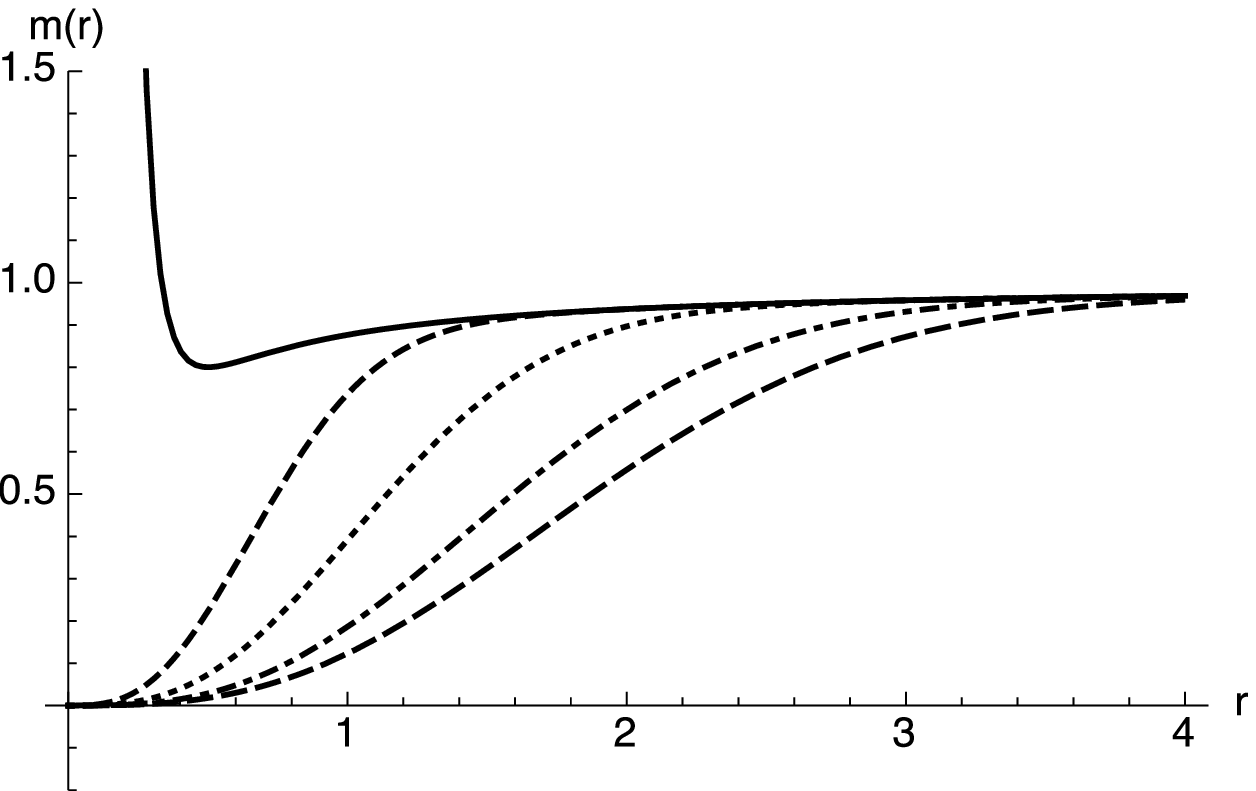}
}
\subfigure[] 
{
    \label{orbits:dd}
    \includegraphics[width=0.4\textwidth]{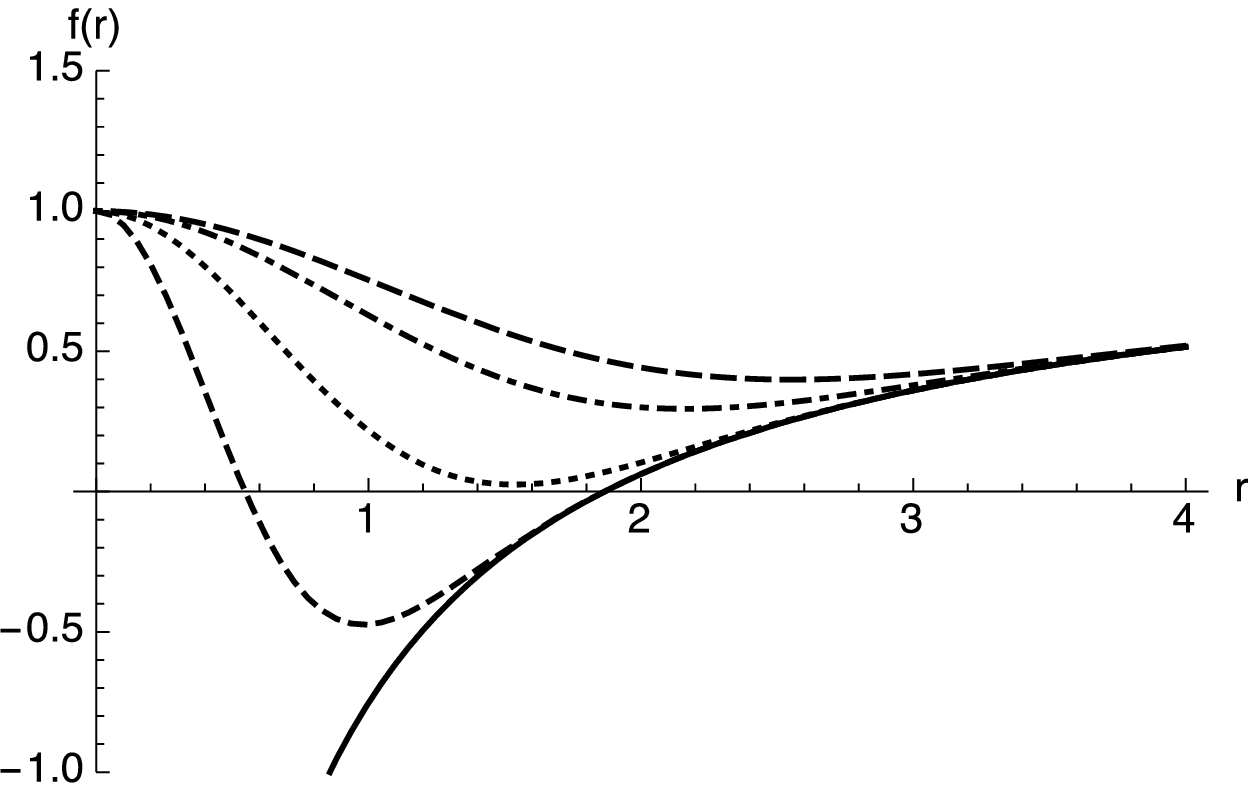}
}
\caption{Plots of the non-commutative functions $m(r)$ and $f(r)$ for values of $\theta = 0.1, 0.25, 0.5$ and $0.7$ (dashed lines from left to right) with $M =1, Q_e = 0.5, A =0$ (first row) and $M=1, Q_e = 0.5, A =1$ (second row). The solid line corresponds to the commutative EEH spacetime.}
\label{fig1} 
\end{figure}

\begin{figure}[h!]
\centering
\subfigure[] 
{
    \label{orbits:ee}
    \includegraphics[width=0.4\textwidth]{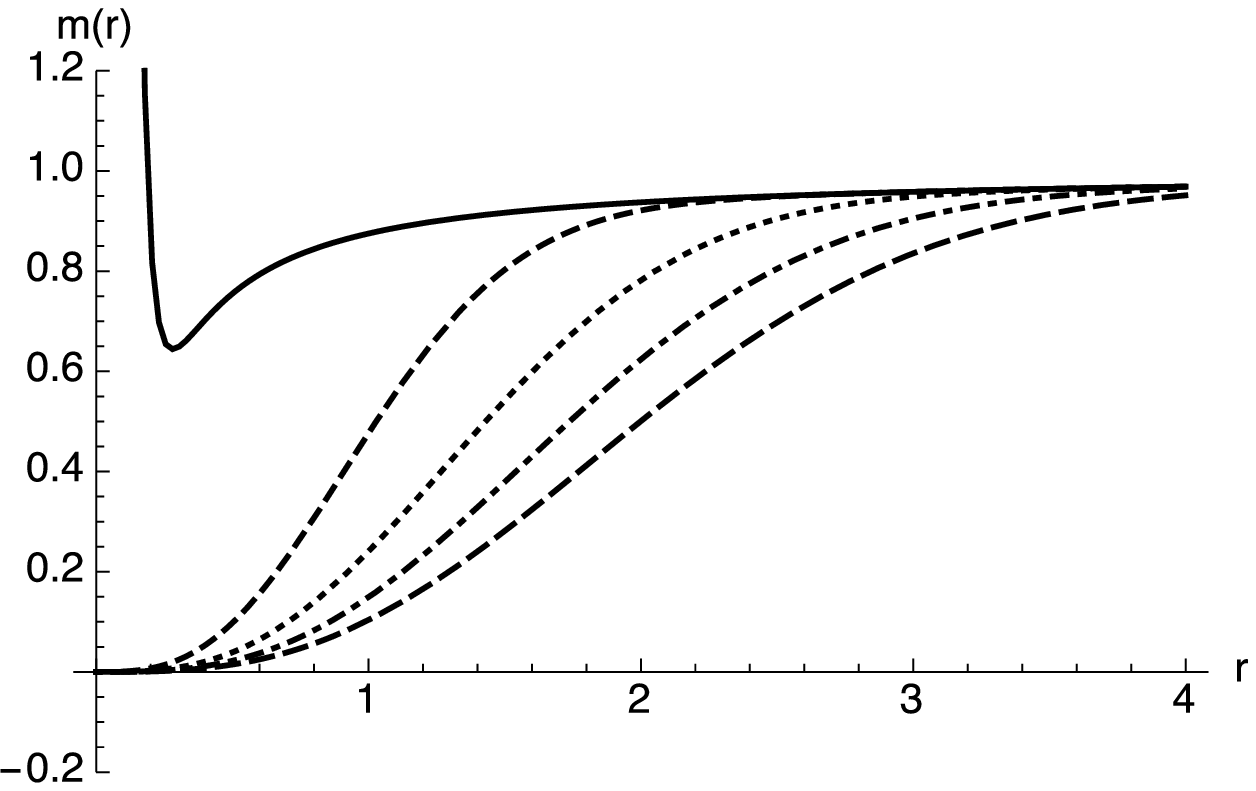}
}
\subfigure[] 
{
    \label{orbits:ff}
    \includegraphics[width=0.4\textwidth]{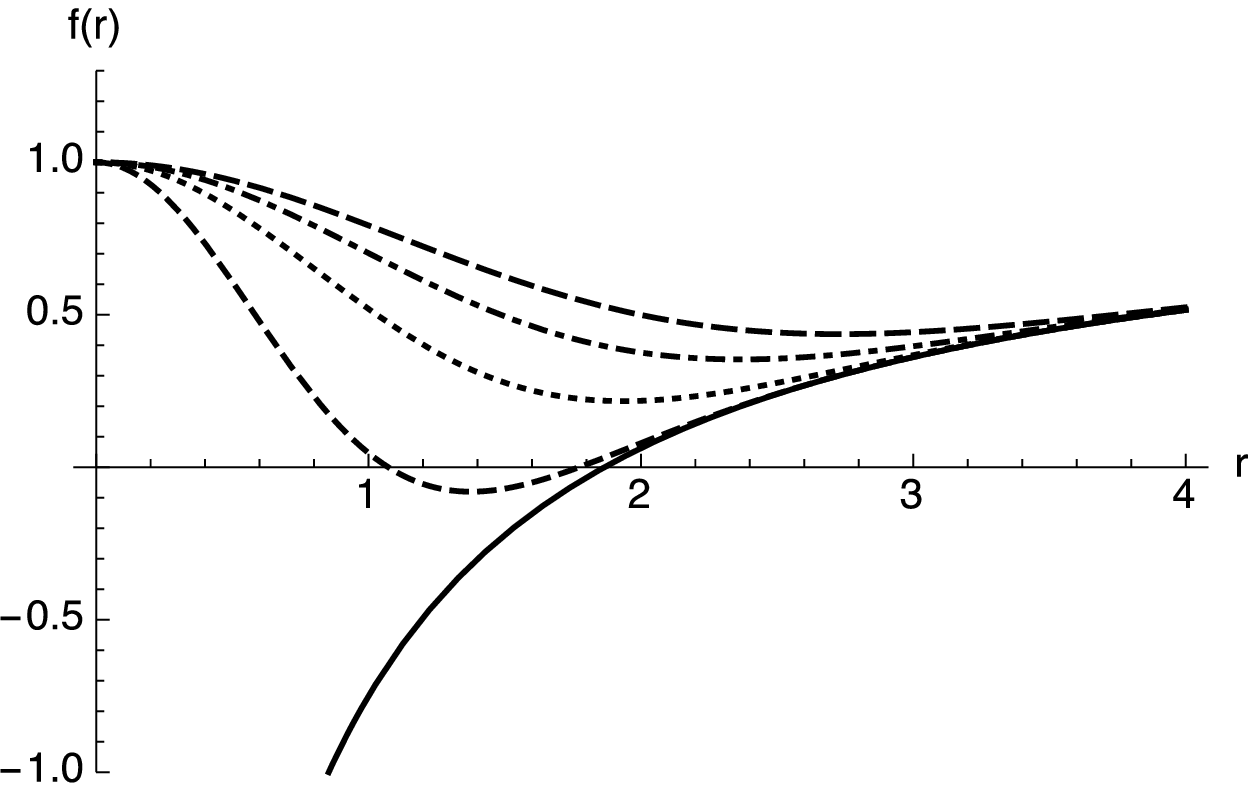}
}
\subfigure[] 
{
    \label{orbits:gg}
    \includegraphics[width=0.4\textwidth]{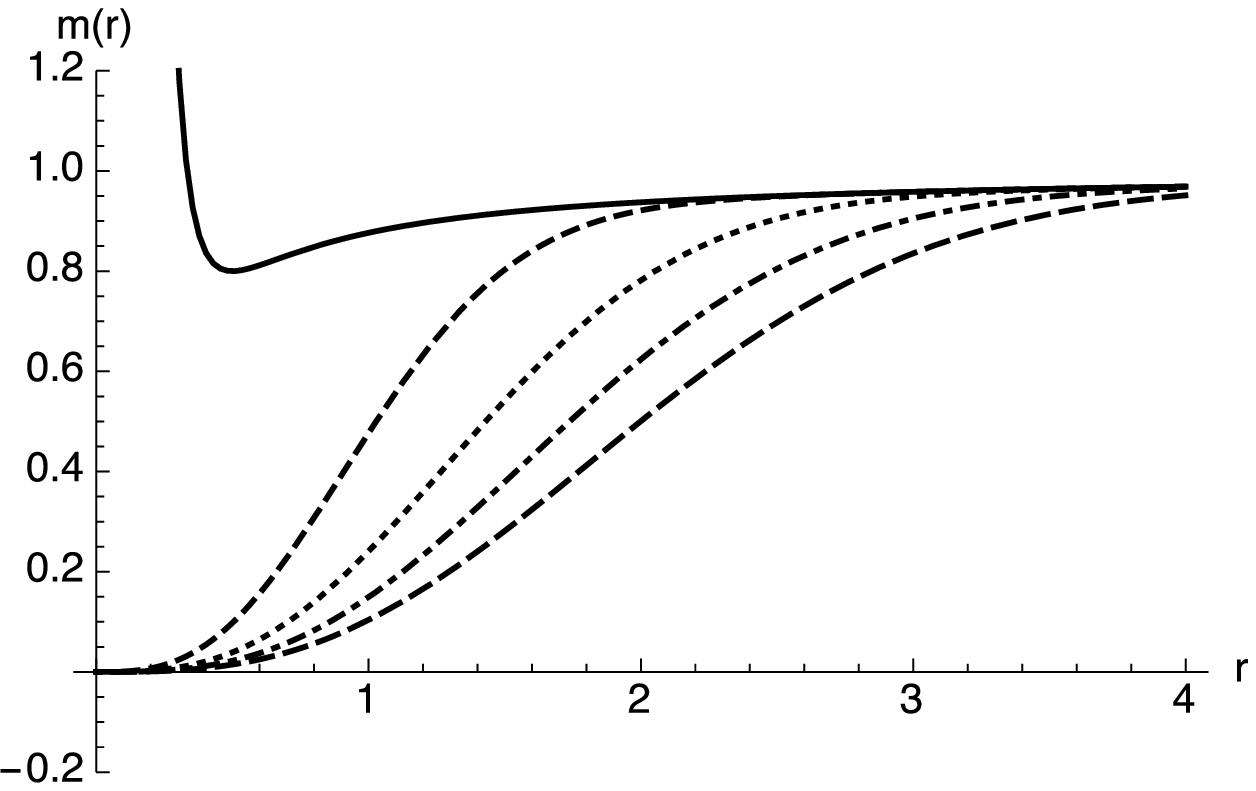}
}
\subfigure[] 
{
    \label{orbits:hh}
    \includegraphics[width=0.4\textwidth]{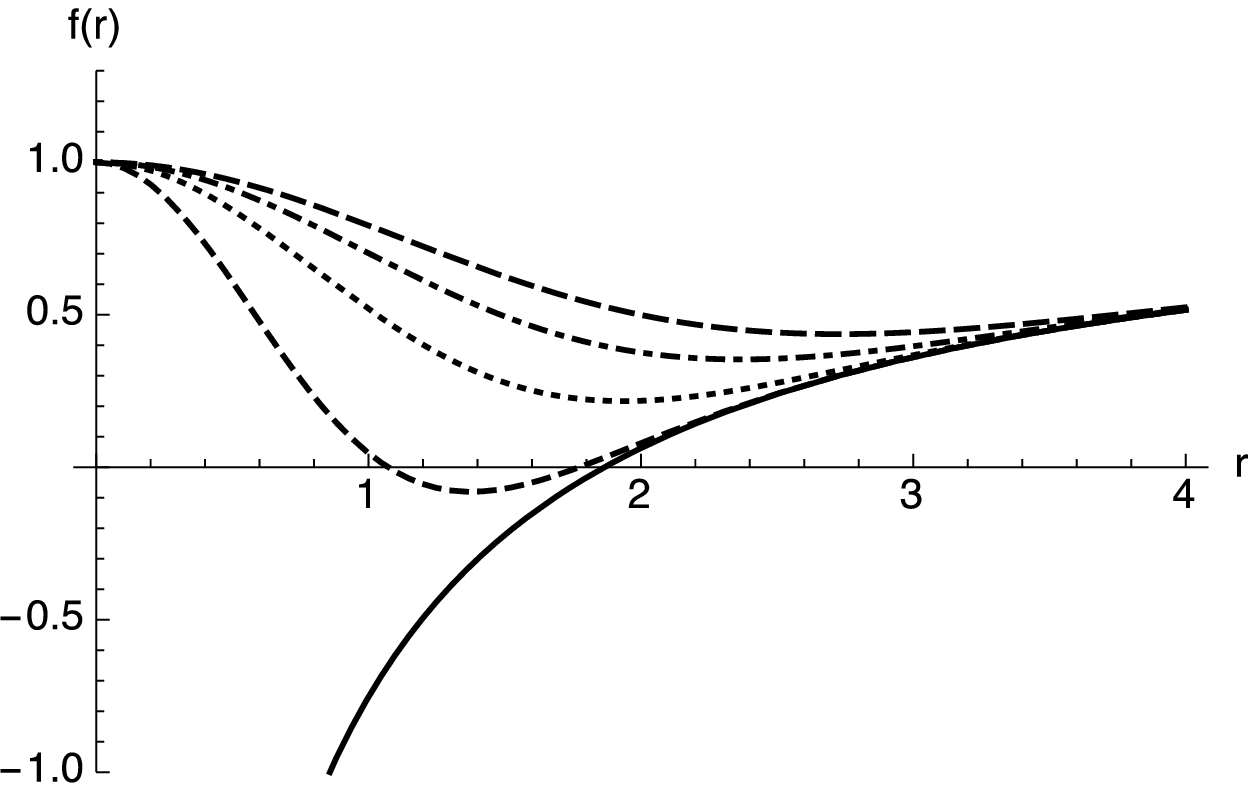}
}
\caption{Plots of the non-commutative functions $m(r)$ and $f(r)$ for values of $\theta = 0.2, 0.4, 0.6$ and $0.8$ (dashed lines from left to right) with $M =1, Q_e = 0.5, A =0.1$ (first row) and $M=1, Q_e = 0.5, A =1$ (second row). The solid line corresponds to the commutative EEH spacetime.}
\label{fig2} 
\end{figure}

For the electric solution, the metric~(\ref{ncecbhadm}) has a horizon determined by condition $f(r_h) = 0$. Due to the complex form of Eq.~(\ref{dsnceeh}), it is clear that an analytic explicit expression for $r_h$ is not possible to obtain due to the presence of the lower incomplete gamma function. In Figs.~\ref{fig1} and~\ref{fig2}, we show the behaviour of the non-commutative functions $m(r)$ and $f(r)$ for different values of the non-commutative parameter $\theta$. As it is seen from these plots, the non-commutative solution is regular at the origin and for values $\theta < \theta_{ext}$ we have a black hole with two horizons. The value $\theta = \theta_{ext}$ is obtained by demanding that the extremal conditions $f(r_h) = 0 = f_{,r} (r_h)$ have a unique solution; the resulting horizon depends on the parameters of the black hole (mass and charge).

If we are interested in the lowest order non-commutative corrections to the classical metric, which are relevant in the analysis of holographic superconductors within the framework of the non-commutative AdS/CFT correspondence~\cite{Pramanik:2015eka}, calculations can be performed to a certain extent as we show now. First, using known identities for $\gamma (a, x)$ and the results in Appendix~\ref{app:a}, we have
\begin{eqnarray}
f(r)&=& 1 - \frac {2M}r + \frac {2M}{\sqrt{\pi\theta}} e^{-r^2/4\theta} + \frac {Q_e^2}{r^2} - 4 \sqrt{ \frac \theta\pi} \frac {Q_e^2}{r^3} e^{-r^2/4\theta}
\nonumber \\[4pt]
&& - \frac {Q_e^2}{\sqrt{2}\pi} \frac 1{\theta} e^{-r^2/4\theta} - \frac A{20} \frac {Q_e^4}{r^6} + 2A \sqrt{ \frac \theta\pi} \frac {Q_e^4}{r^7} e^{-r^2/4\theta}
\nonumber \\[4pt]
&&+ \frac A8 \frac {\alpha Q_e^4}{\pi^{5/2}\theta^3} e^{-r^2/4\theta},
\label{dsefo}
\end{eqnarray}
when $4\theta \ll r^2$. We define now the functions
\begin{eqnarray}
f^c(r) &:=& 1 - \frac {2M}r + \frac {Q_e^2}{r^2} -  \frac A{20} \frac {Q_e^4}{r^6},
\nonumber \\[4pt]
g^\theta (r)&:=& \frac {2M}{\sqrt{\pi\theta}} - 4 \sqrt{ \frac \theta\pi} \frac {Q_e^2}{r^3} - \frac {Q_e^2}{\sqrt{2}\pi} \frac 1{\theta} + 2A \sqrt{ \frac \theta\pi} \frac {Q_e^4}{r^7}
\nonumber \\[4pt]
&&+ \frac A8 \frac {\alpha Q_e^4}{\pi^{5/2}\theta^3},
\nonumber \\[4pt]
f^\theta (r) &:=& g^\theta(r) e^{-r^2/4\theta},
\end{eqnarray}
to be used in the following. If we write the horizon radius $r_+$ in the form $r_+ := r_0 + \alpha$, where $r_0$ is the horizon radius of the commutative case and $\alpha$ contains the non-commutative corrections, we have then that the condition $f (r_+) = 0$ becomes
\be
0 = f^c_{,r} \alpha + \frac 12 f^c_{,rr} \alpha^2 + f^\theta(r_0) + f^\theta_{,r} \alpha + \frac 12 f^\theta_{,rr} \alpha^2,
\ee
up to second order on $\alpha$. Here the first and second derivatives are evaluated at $r = r_0$. Using $f^\theta(r) = g^\theta (r) e^{-r^2/4\theta}$, it follows that
\begin{eqnarray}
f^{\theta}_{,r} &=& \left( g^\theta_{,r} - \frac 1{2\theta} r g^\theta \right) e^{-r^2/4\theta} =:  G_1^\theta e^{-r^2/4\theta},
\nonumber \\[4pt]
f^{\theta}_{,rr} &=& \left( g^\theta_{,rr} - \frac 1\theta r g^\theta_{,r} - \frac 1{2\theta} g^\theta + \frac 1{4\theta^2} r^2 g^\theta \right) e^{-r^2/4\theta}
\nonumber \\[4pt]
&=:& G_2^\theta e^{-r^2/4\theta},
\label{defgs}
\end{eqnarray}
and therefore we arrive to the following quadratic equation on $\alpha$
\be
(1 + b_2 e^{-r_0^2/4\theta}) \alpha^2 + (a_1 + b_1 e^{-r_0^2/4\theta}) \alpha + b_0 e^{-r_0^2/4\theta} = 0,
\ee
where
\begin{eqnarray}
a_1 &:=& 2 \frac {f^c_{,r}(r_0)}{f^c_{,rr}(r_0)},
\nonumber \\[4pt]
b_0 &:=& 2 g^\theta (r_0) / f^c_{,rr}(r_0),
\nonumber \\[4pt]
b_1 &:=& 2 G^\theta_1 (r_0) / f^c_{,rr}(r_0),
\nonumber \\[4pt]
b_2 &:=& G^\theta_2 (r_0) / f^c_{,rr}(r_0).
\end{eqnarray}
If $a_1 \neq 0$, the solution is
\be
\alpha = - \frac {b_0}{a_1} e^{-r_0^2/4\theta} = - \frac {g^\theta(r_0)}{f^c_{,r}(r_0)} e^{-r_0^2/4\theta},
\label{h1}
\ee
where we have used discarded quadratic terms on $e^{-r_0^2/4\theta}$. We find the non-commutative corrections using Eqs.~(\ref{classh1p}),~(\ref{classh1m}),~(\ref{h1}) and
\be
f_{,rr}^c (r) = -\frac {4M}{r^3} + 6 \frac {Q_e^2}{r^4} - \frac {21A}{10} \frac {Q_e^4}{r^8}.
\ee

Eq.~(\ref{h1}) is modified for the case of an extremal classical solution defined by $f^c(r_0) = 0 = f^c_{,r} (r_0)$. In this situation we have $a_1 = 0$ and then
\be
\alpha = \pm \sqrt{-b_0} e^{-r_0^2/8\theta} - \frac 12 b_1 e^{-r_0^2/4\theta}.
\label{he}
\ee
Notice that for this expression to make sense, we need $b_0 \leq 0$. From Eqs.~(\ref{defgs}) we have that the function $G_1^\theta$ is
\begin{eqnarray}
G_1^\theta (r) &=& -14 A \sqrt{\frac \theta\pi } \frac{ Q_e^4}{r^8} - \frac A{\sqrt{\pi\theta}} \frac{Q_e^4}{r^6} - \frac{A \alpha}{16 \pi ^{5/2} \theta ^4} Q_e^4 r - \frac{M r}{\sqrt{\pi } \theta^{3/2}}
   \nonumber \\[4pt]
&& +12 \sqrt{ \frac \theta\pi } \frac{Q_e^2}{r^4} + \frac 2{\sqrt{\theta \pi}} \frac{Q_e^2}{r^2} + \frac 1{2\sqrt{2} \pi  \theta ^2} Q_e^2 r.
\end{eqnarray}
From these results, it is straightforward to calculate the coefficients $b_0$ and $b_1$ from Eq.~(\ref{he}).

\section{Strong, dominant and weak energy conditions}
\label{secc:5}

\begin{figure}[h!]
\centering
\subfigure[] 
{
    \label{fec}
    \includegraphics[width=0.4\textwidth]{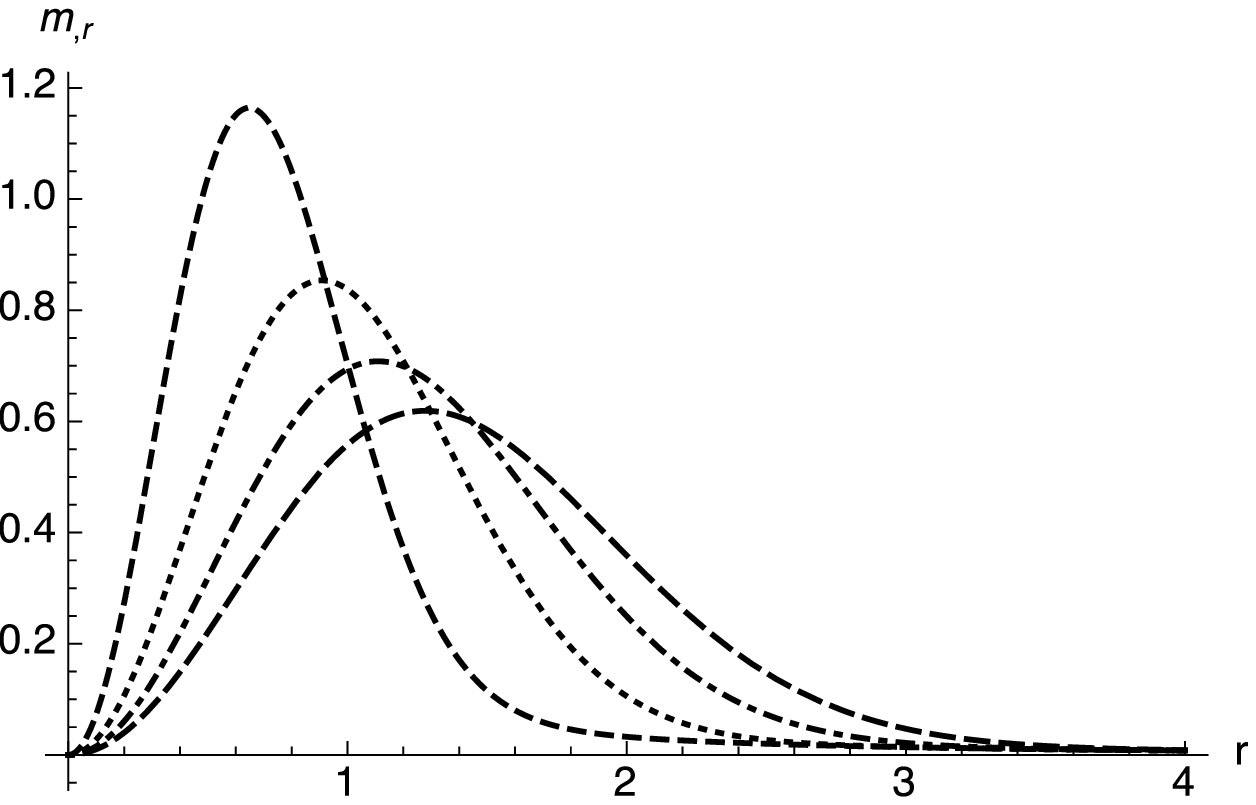}
}
\subfigure[] 
{
    \label{zoomfec}
    \includegraphics[width=0.4\textwidth]{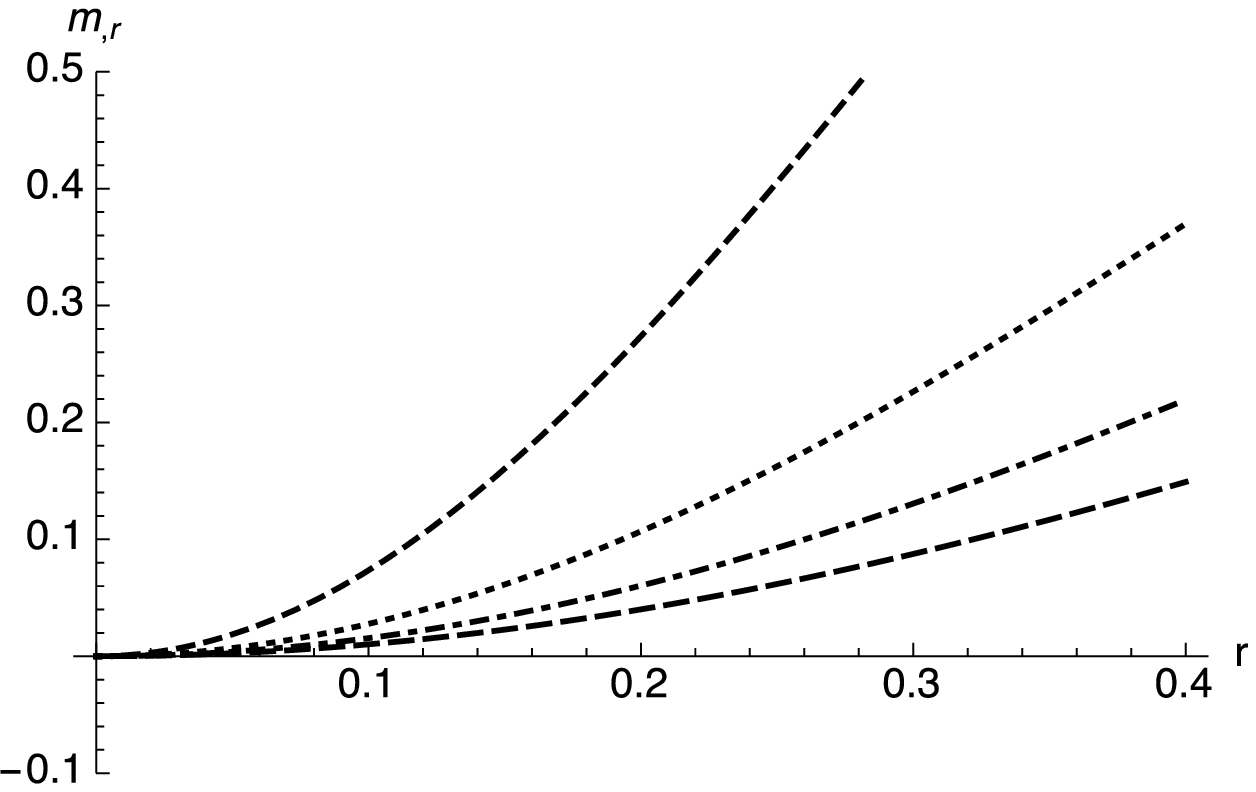}
}
\subfigure[] 
{
    \label{sec}
    \includegraphics[width=0.4\textwidth]{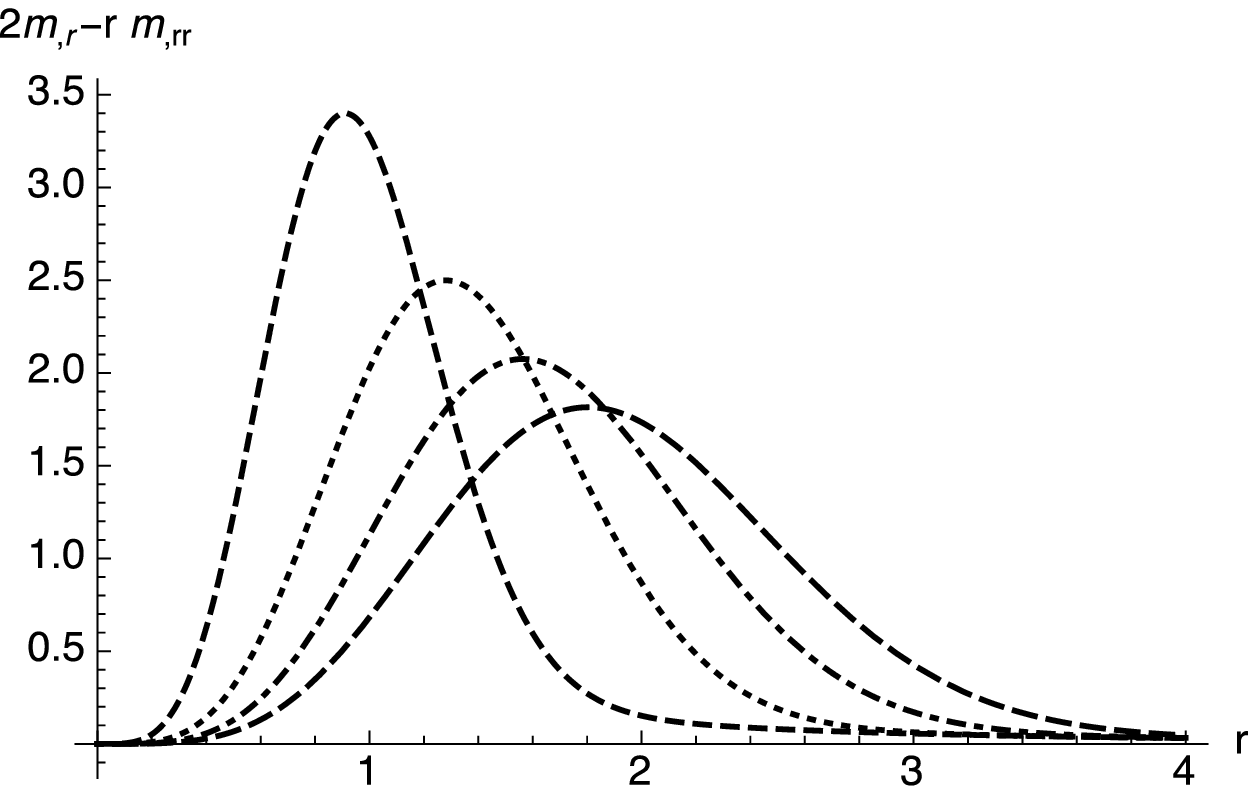}
}
\subfigure[] 
{
    \label{zoomsec}
    \includegraphics[width=0.4\textwidth]{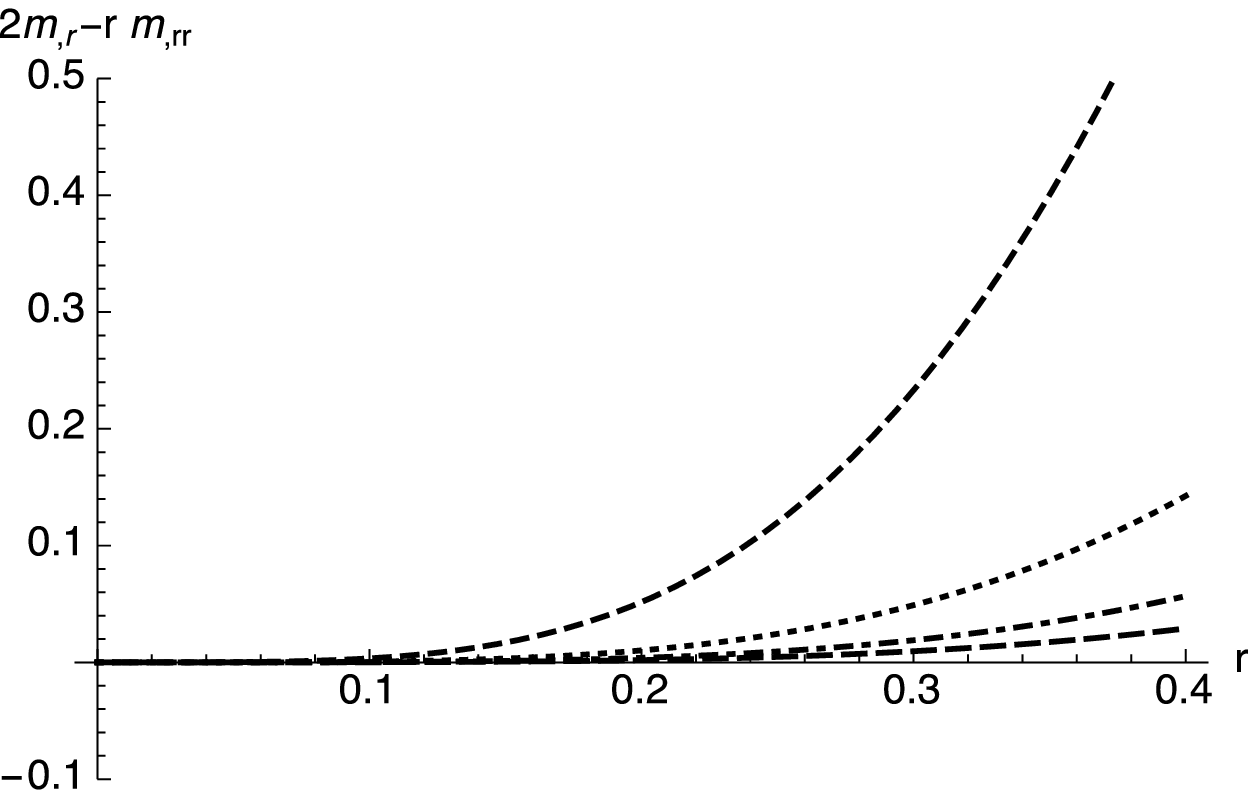}
}
\caption{Plots of the first and second energy conditions for values of $\theta = 0.1, 0.2, 0.3$ and $0.4$ (dashed lines from top to bottom) with $M =1, Q_e = 0.5, A = 0.1$ (EEH black hole with one horizon). The plots on the right are a zoom of the interval $[0, 0.4]$}
\label{fig5} 
\end{figure}

\begin{figure}[h!]
\centering
\subfigure[] 
{
    \label{3hfec}
    \includegraphics[width=0.4\textwidth]{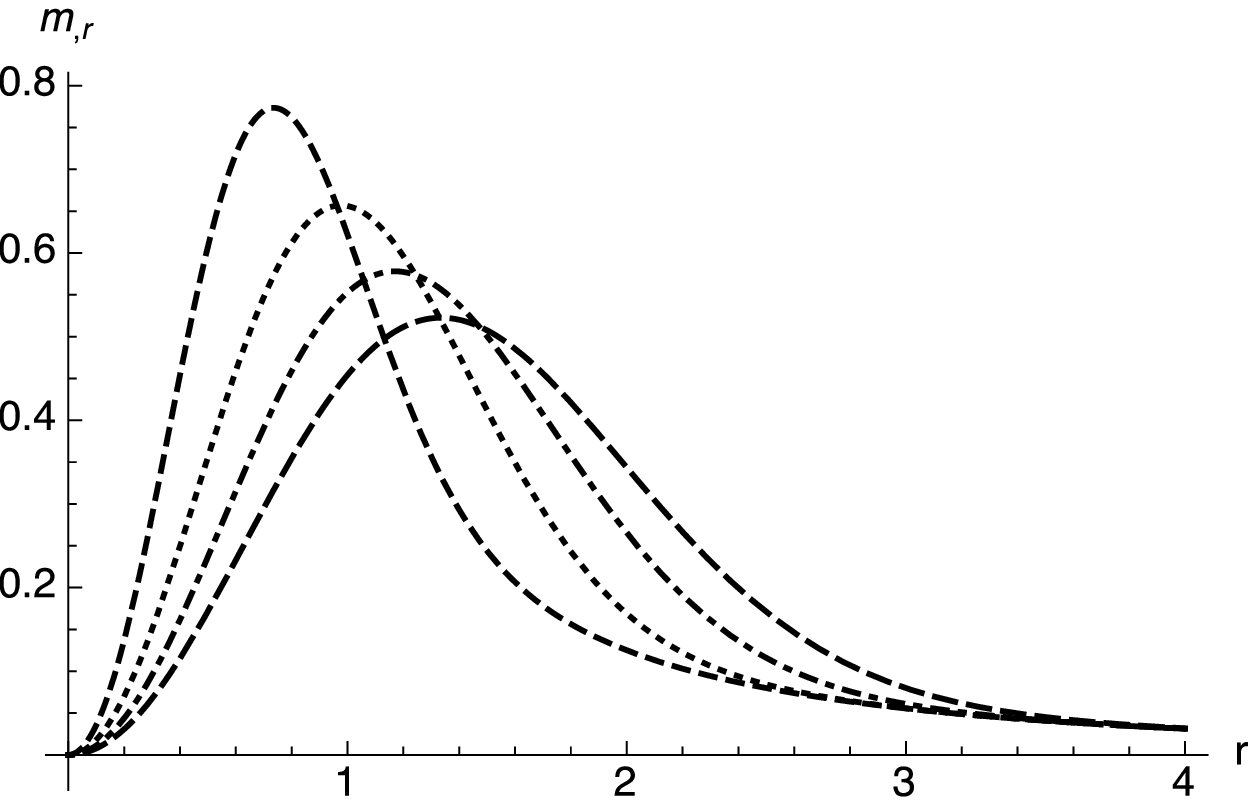}
}
\subfigure[] 
{
    \label{zoom3hfec}
    \includegraphics[width=0.4\textwidth]{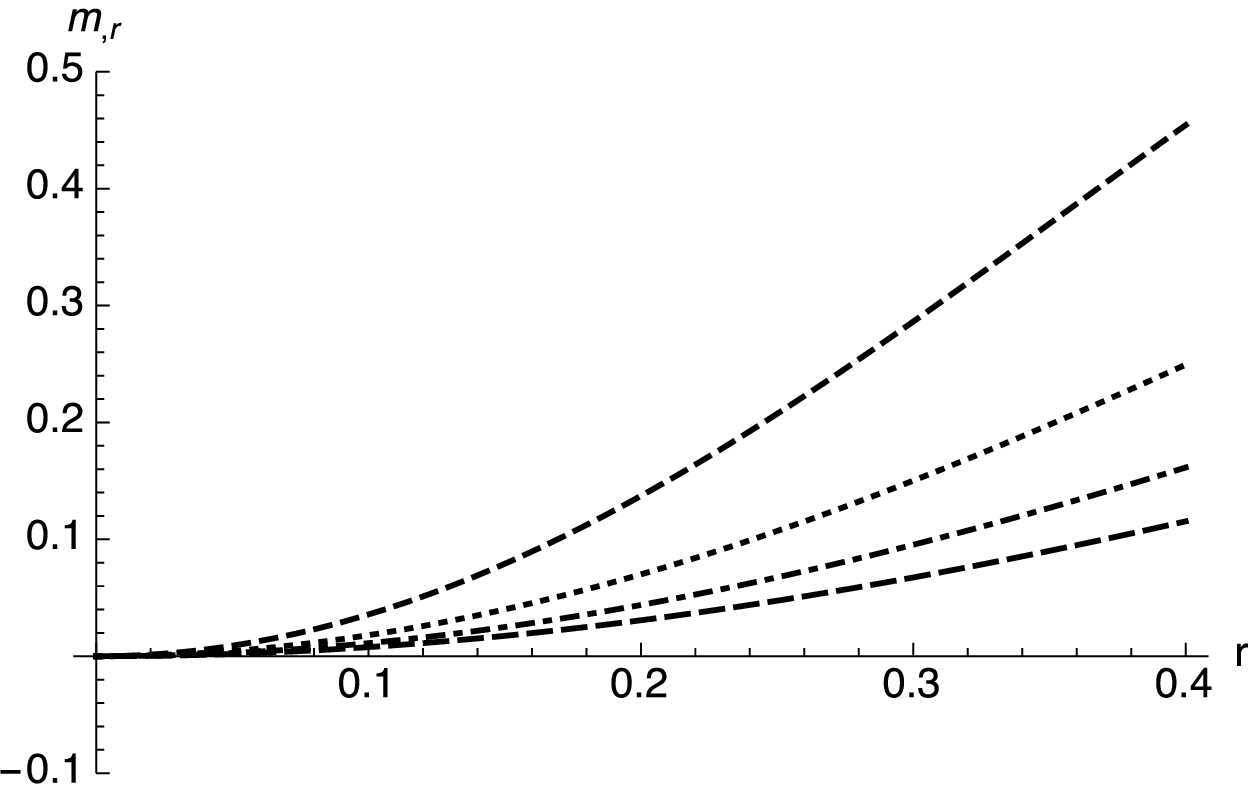}
}
\subfigure[] 
{
    \label{3hsec}
    \includegraphics[width=0.4\textwidth]{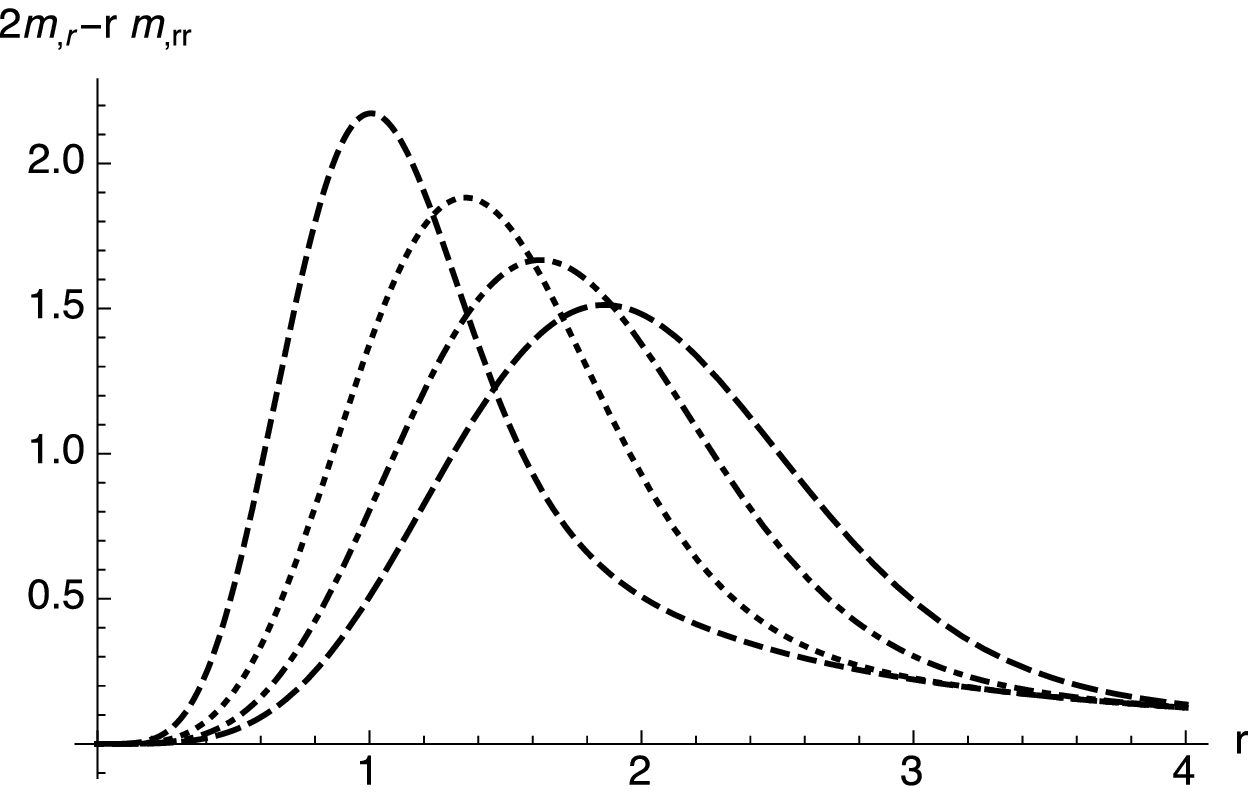}
}
\subfigure[] 
{
    \label{zoom3hsec}
    \includegraphics[width=0.4\textwidth]{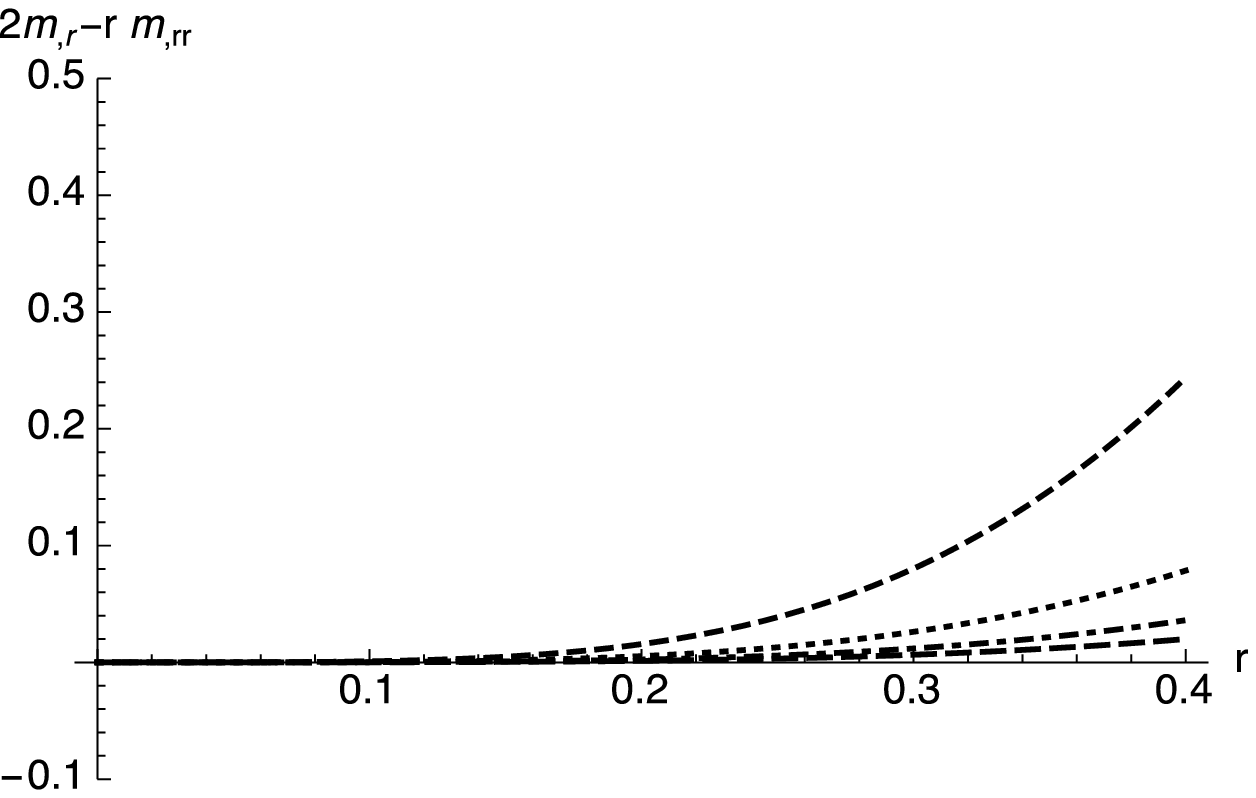}
}
\caption{Plots of the first and second energy conditions for values of $\theta = 0.1, 0.2, 0.3$ and $0.4$ (dashed lines from top to bottom) with $M =1.02, Q_e = 1, A = 0.3$ (EEH black hole with three horizons). The plots on the right are a zoom of the interval $[0, 0.4]$}
\label{fig6} 
\end{figure}

We recall that the strong, dominant and weak energy conditions are
\begin{itemize}
\item Strong Energy Condition (SEC): $T_{\mu\nu} t^\mu t^\nu \geq \frac 12 T^\mu{}_\mu t^\nu t_\nu$ for any timelike vector $t^\mu$,
\item Dominant Energy Condition (DEC): $T_{\mu\nu} t^\mu t^\nu \geq 0$ and $T^{\mu\nu} t_\nu$ must be timelike or null for any timelike vector $t^\mu$,
\item Weak Energy Condition (WEC): $T_{\mu\nu} t^\mu t^\nu \geq 0$, for any timelike vector $t^\mu$.
\end{itemize}
The DEC is often recast as $T^{00} \geq |T^{ij}|$ with $i,j = 1, 2, 3$; it implies the WEC, which is often expressed as the conditions
\be
m_{,r} \geq 0, \qquad 2 m_{,r} \geq r m_{,rr}.
\ee
We consider first the WEC in our model and without loss of generality, we focus on the electric solution. Due to the presence of the lower incomplete gamma function, an explicit expression for the above inequalities is rather cumbersome and not illuminating. For this reason, we show instead in Fig.~\ref{fig5} the corresponding plots for a solution with a single horizon and in Fig.~\ref{fig6} the plots for a solution with three horizons. We remark that in these plots, the two conditions associated to the WEC are satisfied everywhere for the values chosen for the parameters. 

The electric charged EEH solution is regular at the source. It is a known fact that regular solutions violate the SEC, meanwhile the WEC may or may not be satisfied. Let us focus on the behaviour near the origin; a straightforward calculation shows that for $r \to 0$, we have 
\begin{eqnarray}
m_{,r} &=& \frac{\left(A \alpha  Q_e^4 + 16 \pi^2 \theta^{5/2} M - 4 \sqrt{2} \pi ^{3/2} \theta^2 Q_e^2 \right)}{32 \pi^{5/2} \theta^4} r^2 + O(r^4), 
\nonumber \\[4pt]
2m_{,r} - r m_{,rr} &=& \frac{\left(9 A \alpha Q_e^4+144 \pi ^2 \theta^{5/2} M - 4(4+9\sqrt{2}) \pi ^{3/2} \theta^2
   Q_e^2\right)}{576 \pi ^{5/2} \theta ^5} r^4 + O(r^5)
\end{eqnarray}
From these expressions we see then that the non-commutative EEH solution Eq.~(\ref{dsnceeh}), and also its magnetic counterpart, {\it does} satisfy the WEC in a region near the origin depending on the values of the parameters $M, Q_e, A$ and $\theta$; a rough estimate is when $\sqrt{\theta} M \geq 0.263 Q_e^2$. This result is in contrast with the commutative EEH solution where the WEC is always violated near the origin. 

\section{Shadow of the Non-commutative Einstein-Euler-Heisenberg Black Hole}
\label{secc:6}

As astrophysical objects, black holes provide useful insights in the structure of space-time. We now address the formation of a shadow for the non-commutative inspired black holes in EH electrodynamics. From the metric we obtain the Lagrangian
\be
2{\cal L} = -f(r) \dot t^2 + f(r)^{-1} \dot r^2 + r^2 \dot \theta^2 + r^2 \sin^2 \theta \dot \phi^2,
\ee
where a dot denotes derivative with respect to the affine parameter. This Lagrangian provides the starting point to the analysis of orbital motion of test particles and its value determines the orbits under study; for massive and null orbits we have $2{\cal L} = +1, 0$ respectively.

Due to the independence of the metric on the time and azimuthal variable, in general there are two conserved quantities of motion
\begin{eqnarray}
p_t = -f(r) \dot t = - E,
\nonumber \\[4pt]
p_\phi = r^2 \sin^2 \theta \dot \phi = L,
\label{consq}
\end{eqnarray}
related to the energy and angular momentum of the test particle.

In the following we consider the existence of a shadow and therefore restrict ourselves to the case of photon orbits (${\cal L} = 0 $). Solving for $\dot t$ and $\dot \phi$ from Eqs.~(\ref{consq}) and substituting into the Lagrangian, we obtain the equation
\be
\dot r^2 + f r^2 \dot\theta^2 + \frac {L^2 f - E^2 r^2 \sin^2 \theta}{r^2 \sin^2 \theta} = 0.
\ee
Since we are dealing with spherical symmetric solutions to the field equations, the associated shadow of the non-commutative EEH black hole will be circularly symmetric. In consequence, we can fix $\theta = \pi/2, \dot\theta = 0$, knowing that the results are then valid in general. Therefore, we have
\be
\dot r^2 + L^2 \frac f{r^2} - E^2 = 0,
\ee
and from this expression we identify the effective potential
\be
V_{eff} (r) =  L^2 \frac f{r^2} - E^2.
\ee
Circular photon orbits at a radius $r_{ph}$ are then determined by the conditions
\be
V_{eff} (r_{ph}) = 0, \qquad \frac {dV_{eff}}{dr} (r_{ph}) = 0,
\ee
or explicitly
\be
b^2 = \frac {r_{ph}^2}{f(r_{ph})}, \qquad r_{ph} f_{,r} (r_{ph}) - 2 f (r_{ph}) = 0,
\label{impactpar}
\ee
where we have defined the impact parameter $b : = L/E$. In terms of the function $m(r)$, the last equation becomes
\be
r_{ph}[m_{,r} (r_{ph}) +1] - 3m(r_{ph}) = 0.
\label{rph}
\ee

\begin{figure}[h!]
\centering
\subfigure[] 
{
    \label{fig1s}
    \includegraphics[width=0.4\textwidth]{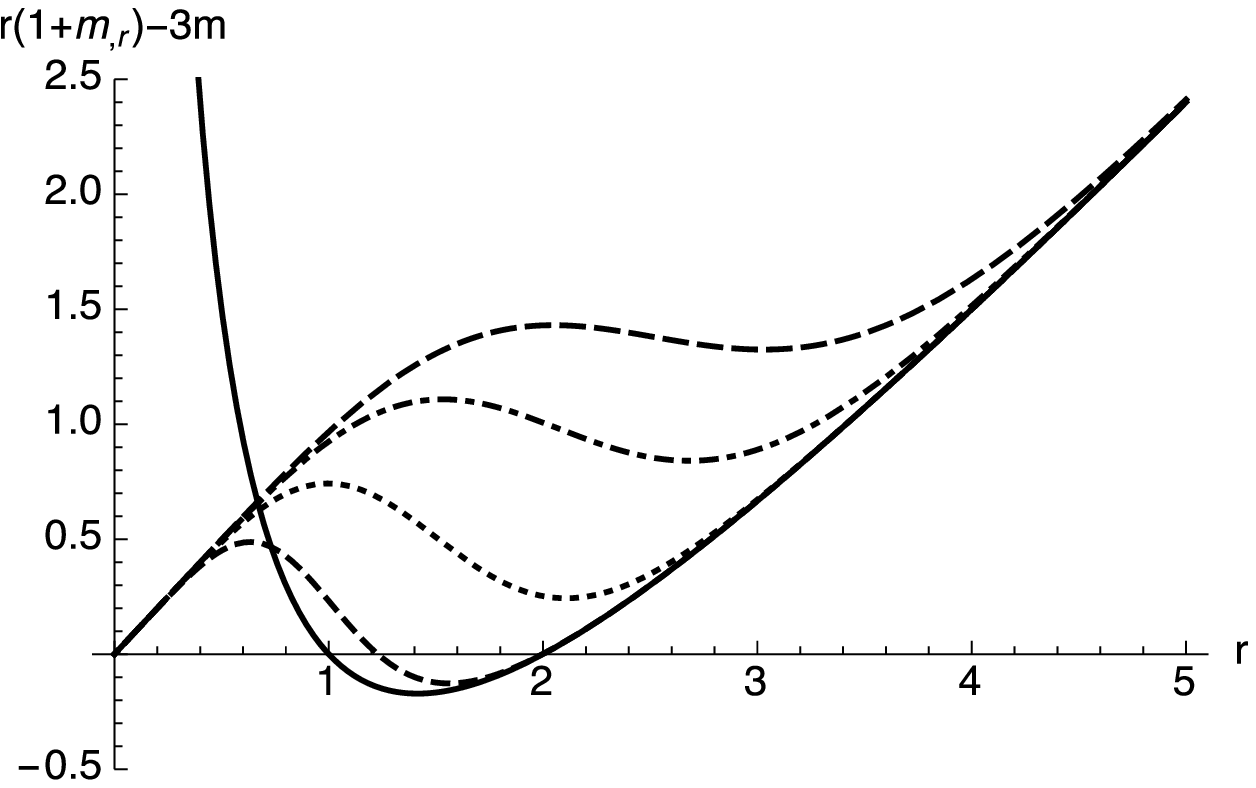}
}
\subfigure[] 
{
    \label{fig2s}
    \includegraphics[width=0.4\textwidth]{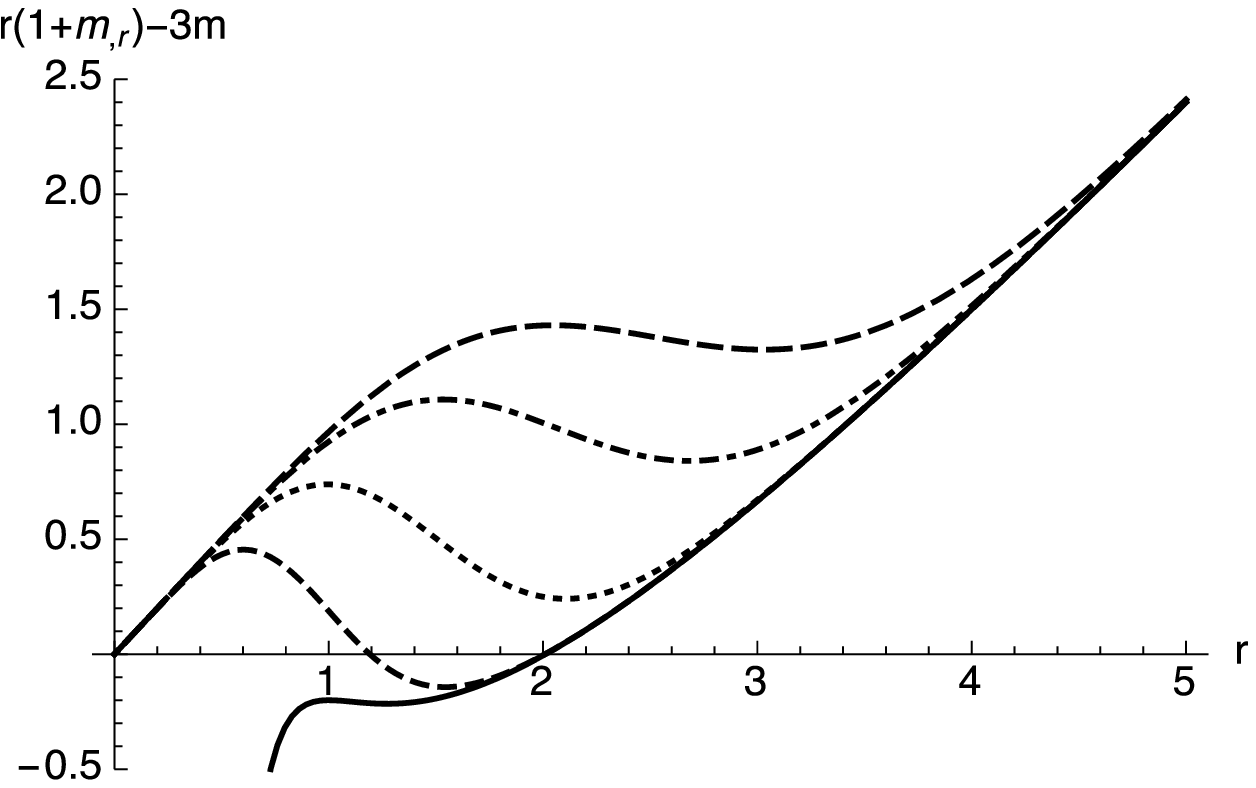}
}
\caption{Plots of the condition determining the existence of circular radial orbits; solid lines correspond to the commutative cases with $M=Q=1$. Fig.~\ref{fig1s} shows the case for $A = 0$ (RN) with $\theta = 0.1, 0.25, 0.5$ and $0.75$ (dashed lines from bottom to top). Fig.~\ref{fig2s} shows the case for $A = 1$ (EEH) with $\theta = 0.1, 0.25, 0.5$ and $0.75$ (from bottom to top). The values of $r$ where the condition vanishes correspond to $r_{ph}$. Notice the existence of a critical value $\theta_{crit}$ above which there are no circular photon orbits.}
\end{figure}

Let us briefly recall the situation for the RN black hole. In this case Eq.~(\ref{rph}) gives the constraint
\be
-3 M + \frac {2Q^2}{r_{ph}} + r_{ph} = 0,
\ee
with solution
\be
r_{\pm ph}^{RN} = \frac{3M \pm \sqrt{9M^2 - 8 Q^2}}2.
\ee
From these values the impact parameter is
\be
(b_+^{RN})^2 = \frac {(3M + \sqrt{9M^2 - 8 Q^2})^4}{8(3M^2 - 2Q^2 + M\sqrt{9M^2 - 8 Q^2})}.
\label{imparrn}
\ee
In Fig.~\ref{fig1s} we show in solid line the left hand side of Eq.~(\ref{rph}) for the commutative RN black hole as a especial case of the non-commutative inspired EH black hole; in this plot we have set $M=Q=1$ (extremal case). For the commutative RN spacetime, there is a circular orbit for photons at $r_{ph} = 1$, at the same location of the extremal horizon, and also at $r_{ph}=2$. Using Eq.~(\ref{imparrn}) we obtain $b^{RN}_+ = 4$; there is always a shadow. The non-commutative inspired RN spacetime ($A=0, M=Q=1$) exhibits a different behaviour: there is a threshold value $\theta_{crit}$ for the non-commutative parameter above which the left-hand side of Eq.~(\ref{rph}) does not vanish, and hence, the non-commutative inspired RN black hole {\it does not} cast a shadow.

\begin{figure}[h!]
\centering
\subfigure[] 
{
    \label{fig1cs}
    \includegraphics[width=0.4\textwidth]{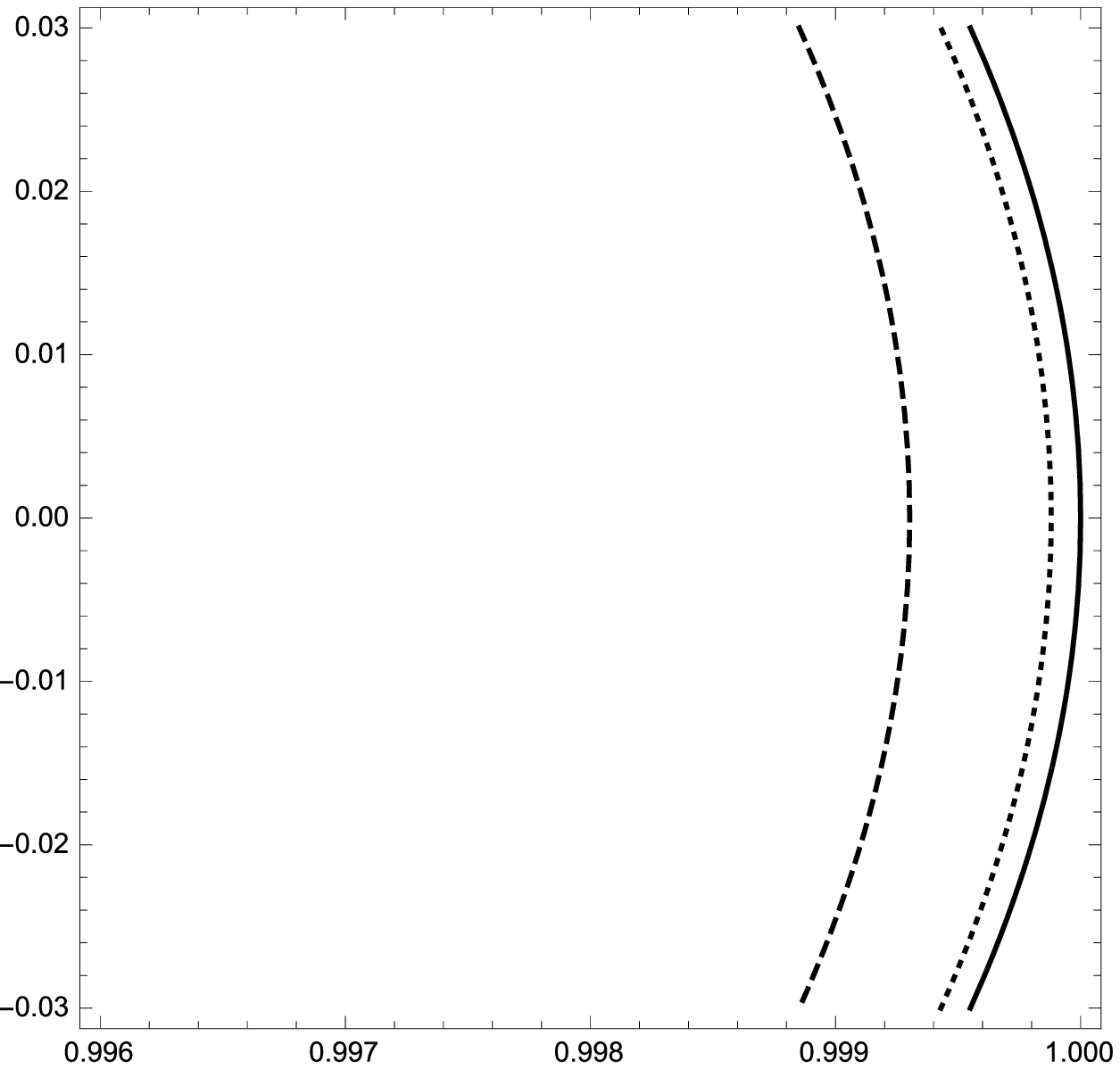}
}
\subfigure[] 
{
    \label{fig2cs}
    \includegraphics[width=0.4\textwidth]{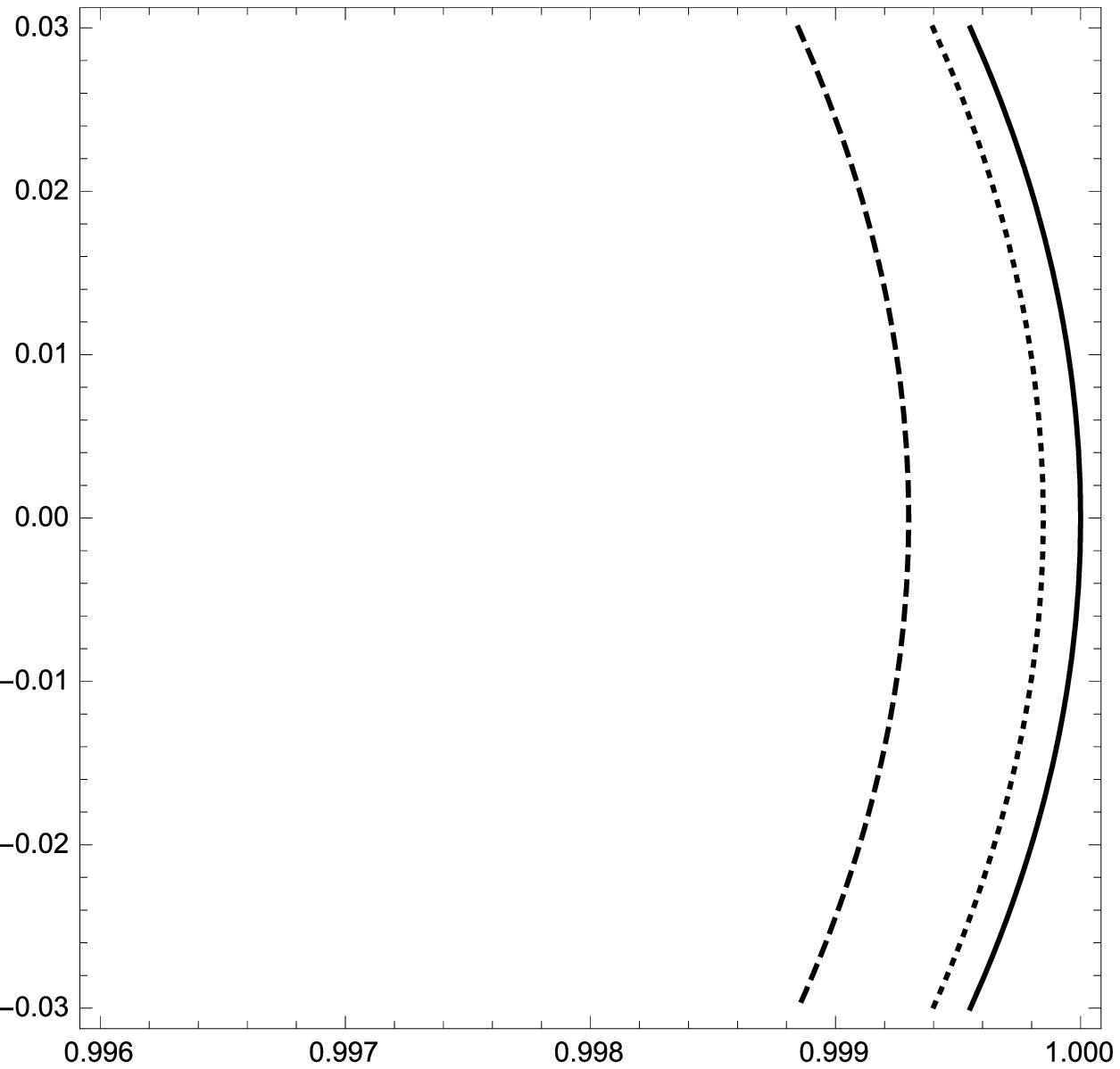}
}
\caption{Detail of the shadow of the non-commutative inspired EEH black hole with $M=Q=1$. Fig.~\ref{fig1cs} shows the case for $A = 0$ (RN) with $\theta = 0, 0.1, 0.12$ (from outer to inner). Fig.~\ref{fig2cs} shows the case for $A = 1$ (EEH) with $\theta = 0, 0.1, 0.12$ (from outer to inner). In each case, the radius of the shadow is given by the ratio $b^{nc}/b^{comm}$ between the non-commutative impact parameter and the commutative value.}
\end{figure}

The situation when $A \neq 0$ is illustrated in Fig.~\ref{fig2s} where we have set $A =1$ and the solid line corresponds to the commutative EEH spacetime; the classical impact parameter is $b^{EEH} = 4.0061$. As in the previous example, there is a value for the non-commutative parameter above which the black hole {\it does not cast} a shadow. We notice also that when there is a shadow, the non-commutative correction to the value $r_{ph}$ seems to be small. Indeed, further analysis shows that the value of $r_{ph}$ for the non-commutative spacetime, either RN or EEH, is less than the commutative one in both situations; the difference in these values is more visible when the EH electrodynamics is turned on. Regarding the impact parameter $b$, we can use Eq.~(\ref{impactpar}) to evaluate it in the non-commutative case. It is seen that we have now a smaller shadow associated with the non-commutative inspired EH black holes. As mentioned before, the correction is small: in Fig.~\ref{fig1cs}, the impact parameter has the values $4, 3.9995$ and $3.9972$ for $\theta = 0, 0.1, 0.12$ (from outer to inner) when $A = 0$ (RN); in Fig.~\ref{fig2cs}, it has the values $4.0061, 4.0055$ and $4.0033$ for $\theta = 0, 0.1, 0.12$ (from outer to inner) when $A = 1$ (EEH).

\section{Conclusions}

We have constructed the non-commutative inspired static electric and magnetic charged black holes coupled to Euler-Heisenberg non-linear electrodynamics. For that purpose, we considered non-commutative smeared distributions that replace the point-like behaviour of sources.

The non-commutative generalisation of the electrical charged EEH spacetime is quite straightforward using the modified field equations. The appearance of lower incomplete gamma functions in the metric is a characteristic feature of these kinds of solutions, providing the mechanism for a non-singular behaviour at the location of the source. The non-commutative inspired magnetic charged EEH spacetime has the same functional form as the non-commutative inspired electrical charged EEH spacetime.

The charged non-commutative inspired EEH metrics show several interesting aspects. They exhibit modifications to the horizon radius that are relevant in connection with the AdS/CFT correspondence and holographic superconductors~\cite{Pramanik:2015eka}. Furthermore, we showed that the weak energy condition is satisfied in a region near the source depending on the values of the parameters $M, Q_e, A$ and $\theta$. We see this last result as a natural consequence of the non-commutative effects that come into play to regularise singularities in the classical solutions.

We also addressed the formation of shadows for the non-commutative inspired EEH metrics. This feature is not present if the non-commutative parameter exceeds a critical value, depending on the mass, charge and EH parameter. When it exists, the non-commutative shadow seems to give small corrections to the classical result; this modification may be susceptible to observation and it is relevant when probing quantum effects in gravity. In this regard, it would be interesting to consider more elaborated non-commutative inspired models where rotation is present.

\acknowledgments

This work was supported by CONACyT Grant A1-S-31056. AM acknowledges support from DAAD.

\appendix
\section{Calculation of integrals involving $\gamma \left( \frac 32, \frac {r^2}{4\theta} \right)$}
\label{app:a}

We illustrate how to calculate the integral
\be
I := \int_0^r \frac {ds}{s^2} \gamma^2 \left( \frac 32, \frac {s^2}{4\theta} \right)
\ee
appearing in the main text. Using the identity~\cite{Abramowitz:1965} $\gamma(\frac a2 + 1, z^2) = \frac a2 \gamma (\frac a2, z^2) - z^a e^{-z^2}$, we have
\begin{eqnarray}
\gamma^2 \left( \frac 32, \frac {s^2}{4\theta} \right) &=& \frac 14 \gamma^2 \left( \frac 12, \frac {s^2}{4\theta} \right) - \frac 1{2\sqrt{\theta}} \gamma \left( \frac 12, \frac {s^2}{4\theta} \right) s e^{-s^2/4\theta}
\nonumber \\[4pt]
&&+ \frac 1{4\theta} s^2 e^{-s^2/2\theta},
\end{eqnarray}
and in consequence
\begin{eqnarray}
I &=& \frac 14 \int_0^r \frac {ds}{s^2} \gamma^2 \left( \frac 12, \frac {s^2}{4\theta} \right)
\nonumber \\[4pt]
&&- \frac 1{2\sqrt{\theta}} \int_0^r \frac {ds}s \gamma \left( \frac 12, \frac {s^2}{4\theta} \right) e^{-s^2/4\theta}
\nonumber \\[4pt]
&&+ \frac 1{4\theta} \int_0^r ds \, e^{-s^2/2\theta}.
\end{eqnarray}
The first integral in the right hand side of the above expression can be rewritten using integration by parts and the identity~\cite{Abramowitz:1965} $\frac {d}{dz} \gamma (\frac a2, z^2) = 2 z^{a-1} e^{-z^2}$; this gives
\begin{eqnarray}
\int_0^r \frac {ds}{s^2} \gamma^2 \left( \frac 12, \frac {s^2}{4\theta} \right) &=& -\frac 1s \gamma^2 \left( \frac 12, \frac {s^2}{4\theta} \right) \Big|_0^r
\nonumber \\[4pt]
&&+ \frac 2{\sqrt{\theta}} \int_0^r \frac {ds}s \gamma \left( \frac 12, \frac {s^2}{4\theta} \right) e^{-s^2/4\theta}
\nonumber \\[4pt]
&=& -\frac 1r \gamma^2 \left( \frac 12, \frac {r^2}{4\theta} \right)
\nonumber \\[4pt]
&&+ \frac 2{\sqrt{\theta}} \int_0^r \frac {ds}s \gamma \left( \frac 12, \frac {s^2}{4\theta} \right) e^{-s^2/4\theta}.
\end{eqnarray}
Therefore, we obtain
\be
I = -\frac 1{4r} \gamma^2 \left( \frac 12, \frac {r^2}{4\theta} \right) + \frac 1{\sqrt{32\theta}} \gamma \left( \frac 12, \frac {r^2}{2\theta} \right).
\ee
Taking the limit $r\to \infty$ we have
\be
\int_0^\infty \frac {ds}{s^2} \gamma^2 \left( \frac 32, \frac {s^2}{4\theta} \right) = \sqrt{\frac {\pi}{32\theta}}.
\ee
This definite integral is used in the calculation of the ADM mass $M$ in Eq.~(\ref{admmass}).

Let us now consider the integral
$$
J := \int_r^\infty \frac {ds}{s^6} \gamma^4 \left( \frac 32, \frac {s^2}{4\theta} \right).
$$
An explicit expression for this integral has not been found. Nevertheless, it can be evaluated in the limit $4\theta \ll r^2$ using the approximation
\be
\gamma \left( \frac a2, x^2 \right) \sim \Gamma\left( \frac a2 \right) - x^{a-2} e^{-x^2},
\ee
valid for large values of $x$~\cite{Abramowitz:1965}. We have then
\begin{eqnarray}
J &\sim& \int_r^\infty \frac {ds}{s^6} \left[ \Gamma^4 \left( \frac 32 \right) - 4 \Gamma^3 \left( \frac 32 \right) \frac s{2\sqrt{\theta}} e^{-s^2/4\theta} \right]
\nonumber \\[4pt]
&=& \frac {\pi^2}{80} \frac 1{r^5} - \frac {\pi^{3/2}}{4\sqrt{\theta}} \int_r^\infty \frac {ds}{s^5} e^{-s^2/4\theta}
\nonumber \\[4pt]
&=& \frac {\pi^2}{80} \frac 1{r^5} - \frac {\pi^{3/2}}{4\sqrt{\theta}} \frac 12 \frac 1{(4\theta)^2} \int_{r^2/4\theta}^\infty du \,u^{-3} e^{-u}
\nonumber \\[4pt]
&=& \frac {\pi^2}{80} \frac 1{r^5} - \frac {\pi^{3/2}}{128\theta^{5/2}} \Gamma \left( -2, \frac {r^2}{4\theta} \right).
\end{eqnarray}
For large values of $x$ we use the approximation $\Gamma(a, x) \sim x^{a-1} e^{-x}$; the above expression is written then as
\begin{eqnarray}
J&\sim& \frac {\pi^2}{80} \frac 1{r^5} - \frac {\pi^{3/2}}{128 \theta^{5/2}} \frac {64\theta^3}{r^6} e^{-r^2/4\theta}
\nonumber \\[4pt]
&=& \frac {\pi^2}{80} \frac 1{r^5} - \frac {\pi^{3/2}}2 \frac {\theta^{1/2}}{r^6} e^{-r^2/4\theta}.
\end{eqnarray}
Using this result we see that the non-commutative EH metric has the form
\begin{eqnarray}
f(z)&=& 1 - \frac {4M}{r\sqrt{\pi}} \left( \frac {\sqrt{\pi}}2 - \frac r{2\sqrt{\theta}} e^{-r^2/4\theta} \right) + \frac 1\pi \frac {Q_e^2}{r^2} \left( \pi - 4 \frac {\sqrt{\pi\theta}}r e^{-r^2/4\theta} \right)
\nonumber \\[4pt]
&& - \frac {Q_e^2}{\sqrt{2}\pi} \frac 1{\theta} e^{-r^2/4\theta} -\frac {4A}{\pi^2} \frac {Q_e^4}r \left[ \frac {\pi^2}{80} \frac 1{r^5} - \frac {\pi^{3/2}}2 \frac {\theta^{1/2}}{r^6} e^{-r^2/4\theta} \right]
\nonumber \\[4pt]
&&+ \frac {A}{8\pi^2} \frac {Q_e^4}r \left[ 1- \frac 2{\sqrt{\pi}} \gamma \left( \frac 32, \frac {r^2}{4\theta} \right) \right] \frac {0.2757}{\theta^{5/2}},
\end{eqnarray}
in the limit $4\theta \ll r^2$.


\end{document}